\documentclass[aip, pop
 amsmath,amssymb,
 reprint,%
]{revtex4-1}

\usepackage{graphicx}% Include figure files
\usepackage{dcolumn}% Align table columns on decimal point
\usepackage{bm}% bold math
\usepackage{comment}
\usepackage[utf8]{inputenc}
\usepackage[T1]{fontenc}
\usepackage{mathptmx}
\usepackage{mathtools}
\usepackage{etoolbox}
\usepackage{xcolor}
\usepackage{nicefrac}
\usepackage{float}

\makeatletter
\def\@email#1#2{%
 \endgroup
 \patchcmd{\titleblock@produce}
  {\frontmatter@RRAPformat}
  {\frontmatter@RRAPformat{\produce@RRAP{*#1\href{mailto:#2}{#2}}}\frontmatter@RRAPformat}
  {}{}
}%
\makeatother
\begin{document}

\preprint{AIP/123-QED}

\title[]{Single-fluid simulation of partially-ionized, non-ideal plasma facilitated by a tabulated equation of state}

\author{G. Su}
\altaffiliation[Now at: ]{Max Planck Institute for Plasma Physics, Boltzmannstr. 2, 85748 Garching b. M., Germany}
\affiliation{ 
Cavendish Laboratory, Department of Physics, University of Cambridge, Cambridge CB3 0HE, United Kingdom
}
\email{george.su@ipp.mpg.de}

% \affiliation{Max Planck Institute for Plasma Physics, Boltzmannstr. 2, 85748 Garching b. M., Germany}

\author{S. T. Millmore}
\affiliation{ 
Cavendish Laboratory, Department of Physics, University of Cambridge, Cambridge CB3 0HE, United Kingdom
}%

\author{X. Zhang}
\affiliation{%
Tokamak Energy Ltd, 173 Brook Drive, Milton Park, Abingdon, United Kingdom
}%

\author{N. Nikiforakis}
\affiliation{ 
Cavendish Laboratory, Department of Physics, University of Cambridge, Cambridge CB3 0HE, United Kingdom
}%

\date{\today}% It is always \today, today,

\begin{abstract}
We present a single-fluid approach for the simulation of partially-ionized plasmas (PIPs) which is designed to capture the non-ideal effects introduced by neutrals while remaining close in computational efficiency to single-fluid MHD. This is achieved using a model which treats the entire partially-ionized plasma as a single mixture, which renders internal ionization/recombination source terms unnecessary as both the charged and neutral species are part of the mixture's conservative system. Instead, the effects of ionization and the differing physics of the species are encapsulated as material properties of the mixture. Furthermore, the differing dynamics between the charged and neutral species is captured using a relative-velocity quantity, which impacts the bulk behavior of the mixture in a manner similar to the treatment of the ion-electron relative-velocity as current in MHD. Unlike fully-ionized plasmas, the species composition of a PIP changes rapidly with its thermodynamic state. This is captured through a look-up table referred to as the tabulated equation of state (TabEoS), which is constructed prior to runtime using empirical physicochemical databases and efficiently provides the ionization fraction and other material properties of the PIP specific to the thermodynamic state of each computational cell. Crucially, the use of TabEoS also allows our approach to self-consistently capture the non-linear feedback cycle between the PIP's macroscopic behavior and the microscopic physics of its internal particles, which is neglected in many fluid simulations of plasmas today. 

\end{abstract}

\maketitle

\section{Introduction} \label{sec: introduction}

Due to the efficiency provided by its single-fluid nature, ideal magnetohydrodynamic (MHD) simulations are the current workhorse for modeling macroscopic plasma behavior \cite{lazerson2016VMEC, anninos2005cosmos++, gonzalez2015solarIdealMHD, fusheng2018lightningIdeal, morley2004HIMAG}. Their popularity is likely to only increase in the coming decades with the growth of the private fusion \cite{greenwald2024preface, chapman2023public, kingham2024TE, creely2023CFS, levitt2023zap, miyazawa2023Helical, mehlhorn2024kms} and aerospace \cite{stone2008nasaCOTS, vozoff2008spacex, bieniawski2022blue_origin, french2022rocketLab} sectors, where stringent limits are placed on both computational resources and time, and the advent of machine-learning-led iterative design processes, where the efficiency of simulations strongly impacts the trade-off between convergence and accuracy \cite{gates2017StellaratorOp, felici2012Actuator, wehner2019transp, piccione2020physicsML}. Despite these advantages, the MHD model suffers from a limited domain of applicability. This is often addressed through the introduction of additional terms into the momentum and energy equations to capture non-ideal effects such as Ohmic resistivity and thermal conduction, resulting in a class of simulation commonly known as "extended-MHD" \cite{hoelzl2021jorek, ferraro2010M3DC1, sovinec2004nimrod, stone2008athena, hansen2021extended,  navarro2017magnus, seyler2011perseus}.

However, extended-MHD simulations are still highly inaccurate for the simulation of low-temperature, partially ionized plasma (PIPs). This is because they fail to capture the additional non-ideal effects introduced by the presence of neutral particles, such as the differing dynamics between the ions and the neutrals, and the impact of the degree of ionization of the plasma on its material properties, including the specific heat ratio and resistivity. Consequently, the modeling of PIPs is currently performed using multi-fluid \cite{ballinger2021uedgeSPARC, bonnin2016SOLPSITER, bufferand2011SOLEDGE, hillier2016slowshock}, or coupled kinetic simulations \cite{reiter2005eirene, ciraolo2019SOLEDGE_EIRENE}, which can be prohibitively high in computational demand for the use case highlighted above. This poses an issue, as the simulation of PIPs is essential to many industrial and academic applications. Examples include modeling the behavior of the edge region in magnetic confinement fusion devices \cite{borodin2022edge, krasheninnikov2020edge}, understanding solar flares and the solar atmosphere \cite{hillier2024partially, heinzel2024partial, soler2024magnetohydrodynamic}, designing lightning protection systems for aircraft \cite{millmore2020lightning, trauble2021improved, wang2023lightning, xiao2023flow_lightning}, and the heat protection and propulsion systems of rockets \cite{auweter1996generators_reentry, panesi2009reentry, macheret2008weakIonize}.

In this paper, we address the need for an efficient yet accurate reduced model of PIPs by presenting a novel single-fluid approach which captures the additional non-ideal effects introduced by neutrals while remaining comparable in efficiency to MHD. Our approach is centered around the description of the entire PIP as a single mixture consisting of electrons, ions and neutrals. This enables us to capture the non-ideal effects introduced by partial-ionization whilst bypassing the need to simulate multiple fluids or having to calculate the source terms related to ionization/recombination through atomic physics simulations\cite{reiter2005eirene, stotler1994degas2}. Instead, the effects of ionization and the differing physics of the mixture's constituent species are encapsulated as contributions to its overall material properties. The practical simulation of this description relies on two fundamental components: an accurate single-fluid model of this partially-ionized mixture, and an efficient method of obtaining its material properties.

The paper is structured as follows. Sec. \ref{sec: theory and model} introduces the new single-fluid model and the tabulated equation of state utilized in our approach, and provides an overview of their derivation. Sec. \ref{sec: numerical methods} describes the numerical methods used. Sec. \ref{sec: results and verification} presents the verification and evaluation of our approach through a range of tests cases specialized for each captured non-ideal effect, and Sec. \ref{sec: conclusion} summarizes the work and introduces areas of future development.

\section{Theory and Model}\label{sec: theory and model}

\subsection{Single-fluid model of partially ionized plasma}

We simulate a model where the PIP is treated as a single mixture of charged and neutral particles. In the absence of external sources/sinks, the density, momentum, and energy of the mixture are conserved, while its material properties change depending on its species composition. The model differs from MHD as it includes an additional evolutionary equation, and additional terms which captures the differing dynamics between the charged and neutral particles - resulting from the neutrals being oblivious of the electromagnetic force - and its impact on the bulk system.

Similar to MHD, the derivation of the model begins with Braginskii's multi-species fluid equations \cite{braginskii1965}. However, we now consider a plasma mixture consisting of neutrals in addition to ions and electrons, resulting in the following three-fluid system:
\\[0.25em]

\emph{Density conservation:}
\begin{gather}
    \frac{\partial \rho_e}{\partial t} + \nabla \cdot (\rho_e \mathbf{v}_e) = 0 \\ 
    \frac{\partial \rho_i}{\partial t} + \nabla \cdot (\rho_i \mathbf{v}_i) = 0 \\ 
    \frac{\partial \rho_n}{\partial t} + \nabla \cdot (\rho_n \mathbf{v}_n) = 0 
\end{gather}

\emph{Momentum balance:}
\begin{gather}
    \frac{\partial \rho_e \textbf{v}_e}{\partial t} + \nabla \cdot (\rho_e \mathbf{v}_e \otimes \mathbf{v}_e) + \nabla p_e + en_e (\mathbf{E} + \mathbf{v}_e \times \mathbf{B}) = \mathbf{R}_e \\
    \frac{\partial \rho_i \textbf{v}_i}{\partial t} + \nabla \cdot (\rho_i \mathbf{v}_i \otimes \mathbf{v}_i) + \nabla p_i - Zen_i (\mathbf{E} + \mathbf{v}_i \times \mathbf{B}) = \mathbf{R}_i \\    
    \frac{\partial \rho_n \textbf{v}_n}{\partial t} + \nabla \cdot (\rho_n \mathbf{v}_n \otimes \mathbf{v}_n) + \nabla p_n = \mathbf{R}_n
\end{gather}

\emph{Energy balance:}
\begin{gather}
    \frac{\partial}{\partial t} ( \zeta_e) + \nabla \cdot \left[ (\zeta_e + p_e) \mathbf{v}_e + \mathbf{q}_e \right] = -e n_e \mathbf{E} \cdot \mathbf{v}_e + \mathbf{R}_e \cdot \mathbf{v}_e + Q_e \\
    \frac{\partial}{\partial t} ( \zeta_i) + \nabla \cdot \left[ (\zeta_i + p_i) \mathbf{v}_i + \mathbf{q}_i \right] = Z e n_i \mathbf{E} \cdot \mathbf{v}_i + \mathbf{R}_i \cdot \mathbf{v}_i + Q_i \\
    \frac{\partial}{\partial t} ( \zeta_n) + \nabla \cdot \left[ (\zeta_n + p_n) \mathbf{v}_n + \mathbf{q}_n \right] = \mathbf{R}_n \cdot \mathbf{v}_n + Q_n
\end{gather}
Here, \(\rho, \mathbf{v}\) and \(p\) are the hydrodynamic variables of density, velocity, and pressure. \(n\) is the number density of particles. \(\zeta\) is the hydrodynamic energy of the fluid given as \(\zeta = \frac{1}{2}\rho v^2 + \rho \varepsilon\). \(\mathbf{q}\) is the conductive heat flux, defined as 
\begin{equation}
    \mathbf{q} = \kappa \nabla T,
\end{equation}
with \(\kappa\) as the thermal conductivity and \(\nabla T\) as the temperature gradient. The subscripts \(i, e, n\) define quantities for the ions, electrons and neutrals respectively. The fluids are coupled to one another through the source terms \(\mathbf{R}\) and \(Q\), which represent the collisional transfer of momentum and energy respectively. 

The three-fluid model is then reduced by combining the electrons and ions into a single fluid in the usual MHD manner \cite{braginskii1965}. The domain of applicability for this assumption is presented in Appendix. \ref{sec: collisionality study}. 

To achieve a single-fluid description, we make the additional assumption that the system is in local thermodynamic equilibrium (LTE), and hence can be described by a single temperature. In this description, the PIP within each computational cell is treated as a single mixture that responds to energy exchanges with its surroundings in a collective manner. Rather than being evolved separately, the effects of the individual species are instead encoded in the material properties of the fluid, which is calculated by summing over contributions from each species weighted by the mixture's composition (elaborated in Sec. \ref{sec: Tabulated EoS}). This treatment allows our simulation to capture the influence of the individual species on the macroscopic plasma behavior despite its single-fluid nature. 

In contrast to temperature, we do \emph{not} assume the equilibriation of the plasma momentum. Therefore, individual ion and neutral velocities are retained in the system and allows for our model to capture the differing dynamics resultant from the fact that the neutrals are oblivious to the magnetic field. Furthermore, it allows for velocity boundary conditions to be set individually for the neutrals and the charged species, which is required for simulations involving the plasma sheath\cite{riemann1991bohm}, or the puffing of neutrals gas into a plasma \cite{stotler2003puffing}. 

To remain consistent with our treatment of the PIP as a single bulk fluid, we perform a change of reference to the center-of-mass frame of the mixture. This results in the (re)definition of the following quantities:

\begin{align*}
\rho = \rho_i + \rho_n, \quad \mathbf{v} = \frac{\rho_i \mathbf{v}_i + \rho_n \mathbf{v_n}}{\rho_i + \rho_n}, \quad \zeta = \zeta_{ei} + \zeta_n, \quad p = p_{ei} + p_n,
\end{align*}
where \(\rho\) and \(p\) are the total density and pressure of the mixture, \(v\) is its bulk flow velocity, and \(\xi_i\), \(\xi_n\) are the mass fraction of ions and neutrals that characterizes its degree of ionization. As usual, the electron mass is assumed to be negligible, giving us \(\xi_n = 1 - \xi_i\). 

In this frame, the ion and neutral velocities are no longer evolved explicitly. Instead, we evolve their relative-motion defined as \(\mathbf{w} = \mathbf{v}_i - \mathbf{v}_n \). This quantity is governed by its own evolutionary equation, and we can retrieve the ion and neutral velocities by combining \(\mathbf{v}\) and \(\mathbf{w}\) as such:
\begin{equation}
    \mathbf{v_i} = \mathbf{v} + \mathbf{\xi_n} \mathbf{w}, \quad \quad \mathbf{v_n} = \mathbf{v} - \mathbf{\xi_i} \mathbf{w}. \label{eqn: retriving i n velocities}
\end{equation}

These assumptions allow us to write the single-fluid model of PIPs as:
\newline

\indent\emph{Density conservation:}
\begin{equation}
  \frac{\partial \rho}{\partial t} + \nabla \cdot (\rho \mathbf{v}) = 0  \label{eqn: PIP single fluid density}
\end{equation}
\newline
\indent\emph{Momentum conservation:}
\begin{equation}
    \frac{\partial \rho \textbf{v}}{\partial t} + \nabla \cdot (\rho \mathbf{v} \otimes \mathbf{v} - \rho \xi_i \xi_n \mathbf{w} \otimes \mathbf{w} + p\mathbb{I}) - \mathbf{J} \times \mathbf{B} = 0 \label{eqn: PIP single fluid momentum}
\end{equation}
\newline
\indent\emph{Energy balance:}
\begin{equation}
    \frac{\partial \zeta}{\partial t} + \nabla \cdot \left[ (\zeta + p) \mathbf{v} + (\zeta + p) \xi_n \mathbf{w} - (\zeta_n + p_n) \mathbf{w} + \mathbf{q} \right] = \mathbf{E} \cdot \mathbf{J} \label{eqn: PIP single fluid energy}
\end{equation}
\newline
\indent\emph{Ion-neutral relative-motion evolution:}
\begin{align}
    \frac{\partial \mathbf{w}}{\partial t} +& (\mathbf{v} \cdot \nabla) \mathbf{w} + (\mathbf{w} \cdot \nabla) \cdot \mathbf{v} 
    + \xi_n (\mathbf{w} \cdot \nabla) \mathbf{w} - (\mathbf{w} \cdot \nabla) \xi_i \mathbf{w} \nonumber \\
    &= \frac{1}{\rho \xi_i} \mathbf{J} \times \mathbf{B} + \frac{\alpha_{en}}{e n_e \rho \xi_i \xi_n} \mathbf{J} -\left(\frac{\nabla p_{ie}}{\rho \xi_i} - \frac{\nabla p_n}{\rho \xi_n}\right) - \frac{\alpha_n}{\rho \xi_i \xi_n} \mathbf{w} \label{eqn: w evolution}
\end{align}
\newline
\indent\emph{Generalized Ohm's law:} 
\begin{equation}
    \mathbf{E} + \mathbf{v} \times \mathbf{B} = \eta_n \mathbf{J} - \frac{\alpha_{en}}{en_e} \mathbf{w} + \frac{1}{en_e} \mathbf{J} \times \mathbf{B} - \xi_n \mathbf{w} \times \mathbf{B} + \frac{\nabla p_e}{en_e} \label{eqn: generalised Ohm's law}
\end{equation}
Here \(\alpha\) is the coefficient of friction between particles species defined as:
\begin{equation}
    \alpha_{ab} = \rho_a \nu_{ab}, \quad \alpha_a = \sum_{\substack{s \neq a}} \alpha_{as},
\end{equation}
where \(\nu_{ab}\) is the collision frequency between species \(a\) and \(b\). \(\eta_{n}\) is the Ohmic resistivity modified by the presence of \(e\)-\(n\) collisions, given as:
\begin{equation}
    \eta_n = \frac{\alpha_{ei} + \alpha_{en}}{e^2n_e^2}.
\end{equation}

We note that the system features partial-pressure (\(p_n\), \(p_{ie}\)) and partial-energy (\( \zeta_n \)) terms. These terms can be problematic for single-fluid models if we have no knowledge of how the bulk pressure and energy of the mixture is distributed among its constituent species. This is where the LTE assumption mentioned above comes in, allowing us to write: 
\begin{equation}
    p = p_e + p_i + p_n = (n_e + n_i + n_n) k_B T, 
\end{equation}
i.e. the partition of pressure is simply dependent on the plasma composition, which is provided by our TabEoS. 

A similar approach can be taken to obtain the partial-energies with slightly more effort. The hydrodynamic energy is defined as:
\begin{equation}
    \zeta_a = \frac{1}{2} \rho_a v_a^2 + \rho_a \varepsilon_a.
\end{equation}
\(m_a\) is known, we can obtain \(v_a\) using \eqref{eqn: retriving i n velocities} and \(\rho_a\) using the plasma composition. Finally, \(\varepsilon_a\) is obtained using \(p_a\) through the closure provided by TabEoS. 

We see that terms involving \(\mathbf{w}\) appear in equations \eqref{eqn: PIP single fluid momentum}, \eqref{eqn: PIP single fluid energy} and \eqref{eqn: generalised Ohm's law}. These terms describe the disruption to the usual electromagnetically driven behavior of the plasma caused by the presence of neutrals. The terms in \eqref{eqn: PIP single fluid momentum} and \eqref{eqn: PIP single fluid energy} describes the relative-motion acting as an additional mechanism for momentum and energy transport, while the terms in \eqref{eqn: generalised Ohm's law} describes modifications to the electromagnetic fields of the plasma as \(\mathbf{w}\) presents an additional movement of charge. 

Let us also briefly focus on \eqref{eqn: w evolution}, the evolutionary equation for \(\mathbf{w}\). The first two RHS terms shows that increased current and field strength leads to the increase of relative-motion. This results directly from the fact that the charged particles experience the electromagnetic force while the neutrals do not, hence increased current or field strength induces greater relative-motion. 

The latter two terms on the RHS of \eqref{eqn: w evolution}, featuring negative signs, are the fluid's efforts to equilibrate. The first of the two describes the convective coupling, where differences in partial pressure lead to equilibrating forces. The final term describes the collisional coupling mentioned previously, where the momentum transfer due to ion-neutral collisions equilibrates the dynamics of the two species. We see that in the absence of these relaxation terms, \(\mathbf{w}\) can grow indefinitely. 

\subsection{Simulated system}
The single fluid system needs to be rewritten to allow for accurate and efficient numerical simulation. The main factors of consideration for this are the system's conservative nature, and its hyperbolicity. Conservative systems allow for discontinuous solutions, leading to the accurate capturing of shock waves  and other physical discontinuities \cite{toro2013riemann}. A hyperbolic system benefits from a more favorable scaling with resolution in regards to the stable time step of the simulation. Under these considerations, we have rewritten our system as:

\begin{equation}
  \frac{\partial \rho}{\partial t} + \nabla \cdot (\rho \mathbf{v}) = 0
  \label{eqn: density eqn sim}
\end{equation}
\begin{equation}
  \frac{\partial \rho \textbf{v}}{\partial t} + \nabla \cdot \left[\rho \mathbf{v} \otimes \mathbf{v} \textcolor{violet}{- \rho \xi_i \xi_n \mathbf{w} \otimes \mathbf{w}} + (p + \frac{1}{2}|B|^2) \mathbb{I} - \mathbf{B} \otimes \mathbf{B} \right] = 0
  \label{eqn: mom eqn sim}
\end{equation}

{\small
\begin{align}
\frac{\partial U}{\partial t} 
&+ \nabla \cdot \Big[ (U + p + \frac{1}{2} B^2) \mathbf{v} 
 + (\mathbf{v} \cdot \mathbf{B}) \mathbf{B} \nonumber \\
& + \textcolor{violet}{(\xi_n \zeta - \zeta_n + \xi_n p - p_n ) \mathbf{w}} \Big] = \textcolor{teal}{\eta_n \left[ \nabla^2 \mathbf{B} \cdot \mathbf{B} + (\nabla \times \mathbf{B})^2 \right]} \label{eqn: energy eqn sim}
\end{align}
}

\begin{equation}
   \frac{\partial \mathbf{B}}{\partial t} + \mathbf{\nabla} \cdot ( \mathbf{B} \otimes \mathbf{v} - \mathbf{v} \otimes \mathbf{B} ) = \textcolor{teal}{\eta_n \nabla^2 \mathbf{B}} \label{eqn: mag induction eqn sim}
\end{equation}

{\small
\begin{align}
    &\textcolor{violet}{\frac{\partial \mathbf{w}}{\partial t} = -(\mathbf{v} \cdot \nabla) \mathbf{w} - (\mathbf{w} \cdot \nabla) \cdot \mathbf{v} - \xi_n(\mathbf{w} \cdot \nabla) \mathbf{w} + (\mathbf{w} \cdot \nabla) \xi_i \mathbf{w}} \nonumber \\
    &\textcolor{violet}{+ \frac{1}{\rho\xi_i} (\nabla \times \mathbf{B}) \times \mathbf{B} + \frac{\alpha_{en}}{e n_e \rho \xi_i \xi_n} \nabla \times \mathbf{B} -\left(\frac{\nabla p_{ie}}{\rho \xi_i} - \frac{\nabla p_n}{\rho \xi_n}\right) - \frac{\alpha_n}{\rho \xi_i \xi_n} \mathbf{w}}. \label{eqn: w evolution sim}
\end{align}
}

We have chosen to retain only the Ohmic term in \eqref{eqn: generalised Ohm's law}, as it enables us to compare our results with a wider range of MHD literature, which is then combined with Faraday's law in the usual MHD manner to arrive at \eqref{eqn: mag induction eqn sim}. We have also chosen to neglect thermal conduction for this initial demonstration of our approach, as its numerical treatment is nearly identical to the treatment of resistive diffusion. 

These modifications create a fully conservative system on the LHS for the fluid density, momentum and energy, without the need for exchange source terms between the different particle species due to ionization/recombination. Diffusive effects have been moved to the RHS to be treated as a source term through operator splitting, which allows our system to accurately capture discontinuities and shocks. We note that energy exchange with the external environment will still come into the system as source terms, such as the loss of energy due to ionization radiation for an optically thin plasma. This energy loss is not currently included in the model and will be addressed in future work.

We have colored the terms related to the ion-neutral relative-motion in purple. We see that in the fully-ionized limit, where \(\xi_i \rightarrow 1\) and \(\xi_n, p_n, \zeta_n \rightarrow 0\), these terms vanish, reducing the system to an extended-MHD model. The terms in blue are from the effects of resistive diffusion, which disappear when \(\eta_n \rightarrow 0\), once more reducing the system, resulting in ideal MHD. This consistency with MHD models enables our approach to efficiently simulate both fully-ionized and partially-ionized plasmas, and allows it to be easily integrated into existing MHD codes.

\subsection{Tabulated Equation of State}\label{sec: Tabulated EoS}

The model still lacks closure, as an expression is required to calculate \(p\) in terms of the evolved variables \(\rho\) and \(\varepsilon\), as well as to obtain the ionization fraction, partial pressure and partial energy terms.  

In MHD simulations, the \(p(\rho,\varepsilon)\) relationship is provided through the ideal gas equation of state:
\begin{equation} \label{eqn: ideal gas EoS}
    p = (\gamma - 1) \rho \varepsilon,
\end{equation}
where \(\gamma\), is the specific heat ratio usually taken as \(\nicefrac{5}{3}\) for a fully ionized plasma, leading to the commonly used relationship of
\begin{equation}
    \frac{3}{2} p = \rho \varepsilon. \label{eqn: common p(e) relation}
\end{equation}
However, this treatment is highly inaccurate for PIPs as the introduction of neutrals greatly changes the physiochemical material properties of the plasma such as its specific heat capacities, and its internal degrees of freedom. These microscopic changes manifest as significant modifications in the macroscopic behavior, including the alteration of the plasma's compressibility (see Appendix \ref{sec: gamma map}), the modification of the MHD wave structure due to changes in \(c_s\), the plasma sound speed, and the modification of the calculation of transport coefficients \(\eta_n\) and \(\kappa\). Consequently, the accurate simulation of a PIP is now highly dependent on the ionization fraction of the plasma which is in turn dependent on its thermodynamic state. 

In our approach, the species composition and the material properties of the PIP are provided through the use of a look-up table. The look-up table serves as a substitute for traditional equations of state and hence will be referred to as the tabulated equation of state (TabEoS). The information is provided specific to the thermodynamic state of each computational cell, allowing our simulation to accurately capture the spatial and temporal variation of material properties with the plasma's evolution. 

\begin{figure}[h] 
\centering    
\includegraphics[width=0.45\textwidth]{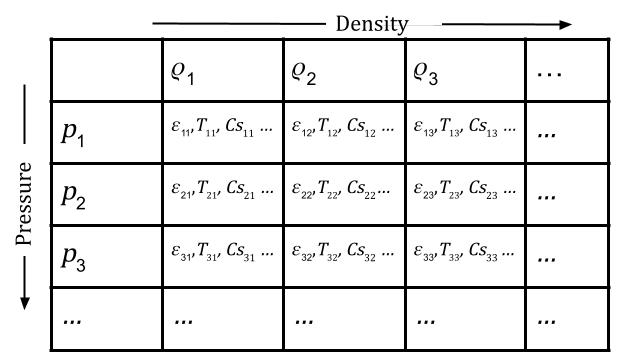}
\caption{Structure of the tabulated equation of state}
\label{fig: tabEoS diagram}
\end{figure}

As shown in Fig.~\ref{fig: tabEoS diagram}, TabEoS is indexed by \(p\) and \(\rho\). For an equilibrium thermodynamic state specified by a particular combination of the two, the following information is currently provided:

\begin{itemize}
    \item Mixture composition:
    \begin{itemize}
        \item Mass fraction of each species, \(\xi_i, \xi_e, \xi_n\)
        \item Molar fraction of each species, \(\chi_i, \chi_e, \chi_n\)
    \end{itemize}
    \item Specific internal energy, \(\varepsilon\)
    \item Mixture temperature, \(T\)
    \item Sounds speed, \(c_s\)
    \item Coefficients of friction, \( \alpha_i \),
    \item Ohmic resistivity, \( \eta_n \)
    \item Thermal conductivity, \( \kappa \).
\end{itemize}

To maximize efficiency, TabEoS is constructed using empirical databases prior to the start of a simulation.
For the results in this work, we have used the \(\text{Mutation}^{++}\) physiochemical library\cite{scoggins2020mutation++}  created by the von Karman Institute for Fluid Dynamics, although many alternatives exist \cite{gordon1994CEA, ern2004eglib, kee1999chemkin}.

In brief, these libraries solve for the chemical equilibrium compositions of ionized gases under a specific thermodynamic state. For Mutation++, this is done using a multiphase equilibrium solver based on the single-phase Gibbs function continuation method \cite{pope2003gibbsCont}. The composition is then used to perform a weighted average over the empirically obtained material properties of pure species to obtain the required properties of the mixture. The empirical data in Mutation++ are provided by the NASA Glenn-coefficent database \cite{mcbride2002nasa, gordon1999thermodynamic}.

\section{Numerical Methods}\label{sec: numerical methods}

Our simulation is discretized using an Eulerian grid. In the interest of accuracy and efficiency, the evolution of our model is divided through operator splitting into three separate updates. Specifically, we follow the Strang-splitting \cite{strang1968construction} formulation:
\begin{equation}
    \mathbf{u}^{k+1} = \mathcal{S}^{(\nicefrac{\Delta t}{2})} \, \mathcal{W}^{(\nicefrac{\Delta t}{2})} \, \mathcal{C}^{(\Delta t)} \,  \mathcal{W}^{(\nicefrac{\Delta t}{2})} \, \mathcal{S}^{(\nicefrac{\Delta t}{2})} \, \mathbf{u}^k
\end{equation}
to achieve second-order accuracy in time. \(\Delta t\) here is the update time step, and the \(\mathcal{C}\), \(\mathcal{S}\) and \(\mathcal{W}\) are the conservative, source, and relative-motion evolution operators respectively. The superscript \(k\) denotes the value of a quantity at time \(t = k \Delta t\).

\subsection{Conservative update (\(\mathcal{C}\))} \label{sec: conservative evolution}

The conservative update evolves the LHS of our model by setting all source terms to zero. This allows us to rewrite the system in the following form:
\begin{equation}
    \frac{\partial \mathbf{u}}{\partial t} + \frac{\partial \mathbf{f}(\mathbf{u})}{\partial x} + \frac{\partial \mathbf{g}(\mathbf{u})}{\partial y} + \frac{\partial \mathbf{h}(\mathbf{u})}{\partial z} = 0.
\end{equation}
where \( \mathbf{u} = (\rho, \rho v_x, \rho v_y, \rho v_z, U, B_x, B_y, B_z, w_x, w_y, w_z )^T\) is our vector of conserved variables (conserved in the context of this update) and \( \mathbf{f}, \, \, \mathbf{g} \) and \( \mathbf{h} \) are the flux functions in each Cartesian direction. \(\mathbf{f(\mathbf{u})}\) is given as
\begin{equation}
    \begin{aligned}
        \mathbf{f(\mathbf{u})} &= 
        \begin{pmatrix}
            \rho v_x \\
            \rho v_x^2 - \rho \xi_i \xi_n w_x^2 + p + \frac{1}{2} |B|^2 - Bx^2 \\
            \rho v_x v_y - \rho \xi_i \xi_n w_x w_y - B_x B_y \\
            \rho v_x v_z - \rho \xi_i \xi_n w_x w_z - B_x B_z \\
            (U + p + \frac{1}{2}|B|^2) v_x + [ \xi_n (\zeta + p) - \zeta_n ] w_x - (\mathbf{v} \cdot \mathbf{B}) B_x \\
            0 \\
            B_y v_x - B_x v_y \\
            B_z v_x - B_x v_z \\
            0 \\
            0 \\
            0 
        \end{pmatrix}
    \end{aligned}
\end{equation}
and similar expressions can be obtained for \( \mathbf{g(\mathbf{u})} \) and \( \mathbf{h(\mathbf{u})} \).

We solve this system using explicit finite volume methods (FVMs), which are chosen for their ability to accurately resolve shocks and other discontinuities \cite{toro2013riemann}. The finite-volume update formula is given as: 
\begin{equation}
\label{eqn:FV update}
\mathbf{u}_i^{k+1} = \mathbf{u}_i^k - \frac{\Delta t}{\Delta x} ( \mathbf{f}_{i+\nicefrac{1}{2}}^k - \mathbf{f}_{i-\nicefrac{1}{2}}^k ),
\end{equation}
where the discretized quantity \( \mathbf{u}_i^k \) is the volume-averaged value of \( \mathbf{u} \) in a cell located at \( x = (i+\frac{1}{2})\Delta x\), while \( \mathbf{f_{i+\nicefrac{1}{2}}^{k}} \) is the flux function between the \(i\)th and \(i+1\)th cell averaged over the \(k\)th time step. The essence of FVMs lies in obtaining accurate estimations of \(\mathbf{f}_{i\pm\nicefrac{1}{2}}\) as it requires information from \(t > t_k\). For our simulation, we have used the slope-limited centered (SLIC) method \cite{toro2013riemann}, which utilizes slope-limited reconstruction of our data to achieve second-order accuracy in both space and time. 

\subsection{Source update (\(\mathcal{S}\))} \label{sec: implicit solver}

The diffusive terms in the system have been extracted from the LHS, and written as a source term. This is because they form a parabolic differential equation of the form:
\begin{equation}
    \frac{\partial Q}{\partial t} = \mathcal{D} \nabla^2 Q, \label{eqn: heat equation}
\end{equation}
where \(\mathcal{D}\) is the diffusion coefficient. Parabolic equations are problematic for explicit numerical methods as their stable time-step scales \(\propto \frac{\mathcal{D}}{\Delta x^2} \), rather than the more favorable \( \propto \frac{1}{\Delta x}\) scaling enjoyed by hyperbolic differential equations \cite{de2013cfl}. Due to the presence of neutrals, resistivity, which acts as the coefficient for the diffusion of the magnetic field, can rise extremely quickly in PIPs (see Appendix. \ref{sec: resis map}), hence rendering the explicit, high-resolution treatment of resistivity inpractical.

In our approach, the diffusive source terms - currently only resistance, but will be extended to include thermal conduction and viscosity in the future - are solved using implicit numerical methods, which are not limited by the CFL condition \cite{de2013cfl}. In particular, we have chosen the Crank-Nicolson method, which is unconditionally numerically stable regardless of the time-step \cite{crankNicolson}. In the case of extreme resistivity, the source update can also be subcycled to achieve physical stability without repeating the conservative update. 

\subsection{Relative-motion update (\(\mathcal{W}\))}

The ion-neutral relative-motion, \(\mathbf{w}\), is not a conserved quantity. Hence, its evolution equation \eqref{eqn: w evolution sim} features non-conservative terms that have also been extracted from the LHS to be treated as a source term. For the results in this paper, they are solved using the explicit Runge-Kutta-2 \cite{butcher2016numerical} method and the second-order centered finite-difference scheme. This combination leads to second-order accuracy in both time and space, which is consistent with the rest of the numerical updates. However, this choice of methods is not integral to our approach and may be substituted for more advanced methods in future work.

\subsection{Divergence cleaning}

Similar to MHD, the simulation of our model is affected by the unphysical growth of \(\nabla \cdot \mathbf{B}\) caused by numerical inaccuracies. For the results in this paper, this is addressed through the use the generalized Lagrangian multiplier-based divergence cleaning method presented by Dedner et al. \cite{Dedner2002}. 

\section{Verification and demonstration} \label{sec: results and verification}
In this section, we validate the accuracy of our approach and demonstrate its ability to capture the non-ideal affects introduced by partial-ionization and their dependence on the plasma's thermodynamic state.

This is achieved by initializing standard MHD test cases in different thermodynamic environments. By rescaling the test parameters accordingly, we capture only the impact of the changes to the material properties of the plasma. Consequently, MHD simulations using an ideal gas EoS which do not capture these changes will exhibit identical results regardless of the initial state. 

The initial state for a general test cases can be written as:
\begin{equation*}
\mathbf{u}_0 = (\rho, v_x, v_y, v_z, p, B_x, B_y, B_z)^T.
\end{equation*}
\(w_x\), \(w_y\) and \(w_z\) are not included as they are always initialized to 0. The thermodynamic state is adjusted through the density and pressure scale factors \(\beta_{\rho}\) and \(\beta_p\). The rest of the variables are then rescaled accordingly through the rescaling vector \(\mathbf{S}\):
\begin{equation*}
\mathbf{S} = (\beta_\rho, \frac{\sqrt{\beta_p}}{\beta_\rho}, \frac{\sqrt{\beta_p}}{\beta_\rho}, \frac{\sqrt{\beta_p}}{\beta_\rho}, \beta_p, \sqrt{\beta_p}, \sqrt{\beta_p}, \sqrt{\beta_p})^T.
\end{equation*}
for the simulated quantities and the rescaling factor \(\nicefrac{\beta_\rho}{\beta_p}\) for time. Altogether, the actual initialized state \(\mathbf{u}^*_0\) and final time \(t_f^*\) of the simulation are given as:
\begin{equation*}
\mathbf{u}^*_0 = \mathbf{u}_0 \cdot \mathbf{S}, \quad \quad t_f^* = \frac{\beta_\rho}{\beta_p} t_f.
\end{equation*}

As the relationship between the plasma's thermodynamic state and its material properties are highly non-linear in nature, the expected behavior of a test case is not always obvious. To address this, we have conducted a systematic study of how the material properties indicative of specific non-ideal effects varies with the thermodynamic state of plasma, which is presented in Appendix. \ref{sec: PIR mapping}.

Numerically, all the presented results were simulated using the SLIC scheme with a CFL number of 0.7.

\subsection{Plasma compressibility}

\begin{figure}[h] 
\centering    
\includegraphics[width=0.5\textwidth]{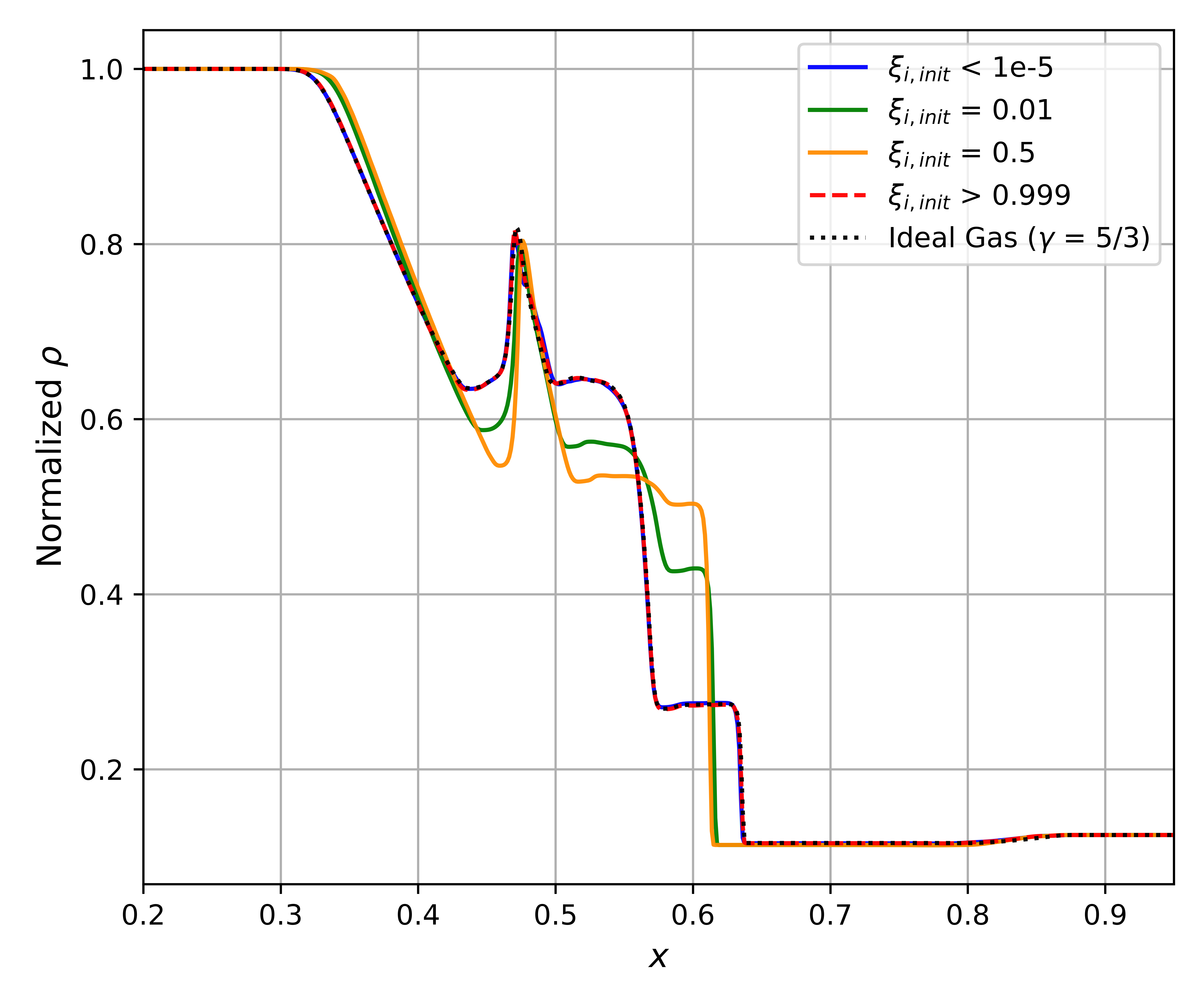}
\caption{Comparison of the \(t = 0.1\) normalized density profiles of the Brio Wu test simulated using TabEoS for different initial ionization fractions. A result simulated using the ideal gas EoS, which does not depend on \(\xi_i\), is also provided as reference. The TabEoS results shows the compressibility of the plasma varying with \(\xi_i\), exemplified by the changes to the jump in density and temperature in the post-shock region. The variation is non-monotonic in nature, where the maximum deviation between the TabEoS and ideal gas results is observed at \(\xi_{i, init} = 0.5\), matching the predictions from Appendix \ref{sec: gamma map}.}
\label{fig: BW gamma study}
\end{figure}

We begin by showing the effect of closing our system using TabEoS instead of the ideal gas EoS \eqref{eqn: ideal gas EoS}. Fig.~\ref{fig: BW gamma study} shows a comparison of results from the well-known Brio-Wu shock tube test \cite{brioWu1988}:
\begin{gather*}
\mathbf{u}_0 =
\begin{cases}
    (1,0,0,0,1,0.75,1,0)^T & x \leq 1/2 \\
    (0.125,0,0,0,1,0.75,-1,0)^T & x > 1/2 \\
\end{cases} \\
 x \in [0,1], \quad t_f = 0.1
\end{gather*}
initialized in thermodynamic states with different degrees of ionization. The boundary conditions for the test are transmissive, and all other non-ideal effects have been switched off. 

We observe that the changes in the ionization fraction of the plasma lead to a significant modification of its compressibility, as showcased by changes in the jump in density and temperature in the post-shock region. 

The trend of this modification is non-monotonic in nature. The results from both the highly ionized and neutral limits resembling those simulated using an ideal gas EoS. This is expected since in these limits, the PIP physiochemically resembles a monoatomic ideal gas, and few ionization events take place. 

As the PIP becomes more mixed, we observe a greater jump in its density in the post-shock region, indicating an increase of its compressibility. Furthermore, we also see that the wave positions for the results initialized with \(\xi_{i, init} = 0.01\) and 0.5 are closer to the initial interface than those of the cases initialized in the neutral and fully-ionized limits, suggesting a slower normalized sound speed (normalized by density and pressure) for the more mixed plasma, which is usually associated with greater compressibility. This is also consistent with the reduction in sound speed observed in other fluid mixtures\cite{farmer2021_woodspeed, kieffer1977sound}.

These results demonstrate our approach's ability to capture the dramatic impact the changes in material properties can have on the macroscopic behavior of the plasma, and also highlights the inaccuracy of the ideal gas EoS for simulating partially-ionized phenomena.

\subsection{Ion-neutral relative-motion}
Next we demonstrate our simulation's ability to capture differing dynamics between the charged particles and the neutrals. We will utilize the slow-mode shock test presented by Hillier, Takasao and Nakamura \cite{hillier2016slowshock}. The test is initialized as:

\begin{gather*}
\mathbf{u}_0 =
\begin{cases}
    (1,0,0,0,1,0.9, 1,0)^T & x < 0 \\
    (1,0,0,0,1,0.9, -1,0)^T & x \geq 0 \\
\end{cases}
\end{gather*}
where the \(x < 0\) state is not actually simulated but is captured by a reflective boundary condition at \(x=0\). The rightward boundary condition is set as transmissive. The jump in \(B_y\) at \(x = 0\) induces a fast rarefaction wave, followed by a slow shock wave. Since both of these waves are electromagnetically driven, they only directly affect the charged particles, causing the neutrals and ions to exhibit differing dynamics in the absence of coupling. 

The initial state is kept constant for this test, with \(\alpha_{\rho} = 0.65 \times 10^{-7} \text{kg} 
\cdot \text{m}^{-3} \), and \(\beta_p = 10\) Pa, leading to an ionization fraction of \(\sim\)0.15. Instead, we vary the final time of the simulation to compare the degree of coupling between the ions and the neutrals for phenomena of different timescales. The \(x\)-axis has been normalized by the final time to facilitate this comparison. 

\begin{figure}[] 
\centering    
\includegraphics[width=0.5\textwidth]{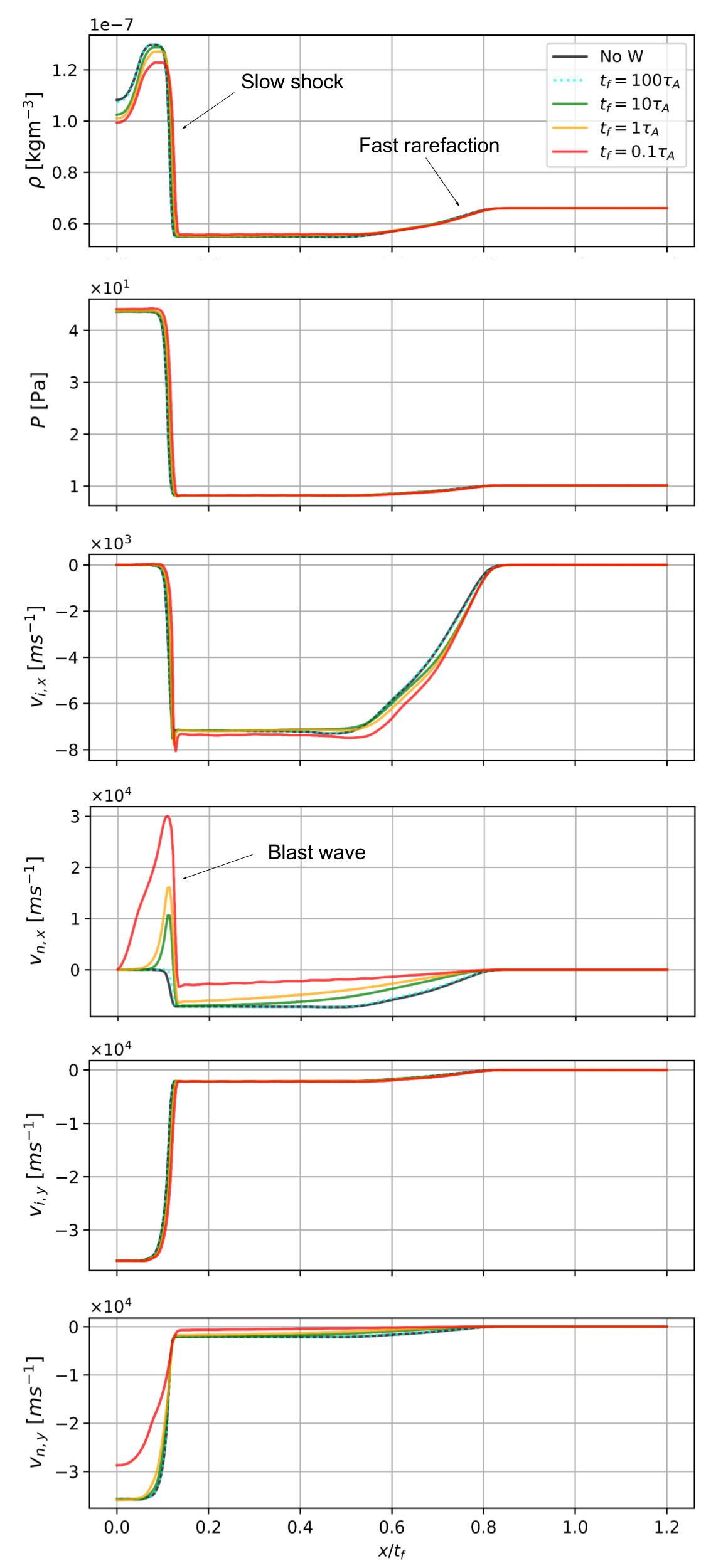}
\caption{Density, pressure, ion and neutral velocity profiles for five runs of the slow-mode shock test initialized with \(\xi_i \sim 0.15\). The ion-neutral relative-motion is turned off for the "no-W" run while the others are run with different final times (\(t_f\), chosen as multiples of the Alfvén time), hence changing the time available for collisional coupling. The \(x\)-axis has been normalized by \(t_f\) to enable comparison. Large differences between the ion and neutral dynamics are observed at low \(t_f\), which affects the overall plasma density and pressure, but are diminished with the increase of \(t_f\).}

\label{fig: slow shock 1}
\end{figure}

Fig.~\ref{fig: slow shock 1} shows the results for final times of \(0.1 \tau_A\), \(\tau_A\), \(10\tau_A\) and \(100\tau_A\) - where \(\tau_A\) is the Alfvén time of the bulk plasma calculated using the initial density. 

Our results agree with the two-fluid simulations (neutral and charged) by Hillier et al.\cite{hillier2016slowshock}. We see that in the short timescale cases, in the region following the rarefaction, a significant velocity has only been induced in the ions, while \(v_{n,x}\) and \(v_{n,y}\) remain close to zero. However, since both species feature a common temperature, the slow shock also leads to the heating of the neutrals, and the creation of a steep pressure gradient in the \(x\)-direction. Since the neutrals in the pre-shock region are almost stationary, this highly localized heating leads to an explosive blast wave \cite{needham2010blast}, characterized by a dramatic but localized jump in \( v_{n,x} \). These results show that for a pressure of \(p = 10 \) Pa, the charged and neutral dynamics are not equilibrated at the Alfvénic timescale, and hence cannot be simulated using only a MHD simulation, but can however be captured by our model. 

As timescale is increased, more opportunities are created for the collisional transfer of momentum between the particle species. This leads to the neutrals being dragged along by the charged particles, evidenced by the increase in both \(v_{n,x}\) and \(v_{n,y}\) in the post-rarefaction region. The increase in neutral velocity reduces the localization of temperature in the post-shock region, reducing the strength of the blast wave. We see that the profiles for the \(t_f = 100\tau_A\) case completely matches the MHD simulation, indicating that the two species are completely equilibrated at this timescale.

\subsection{Resistivity}

\begin{figure}[t] 
\centering    
\includegraphics[width=0.4\textwidth]{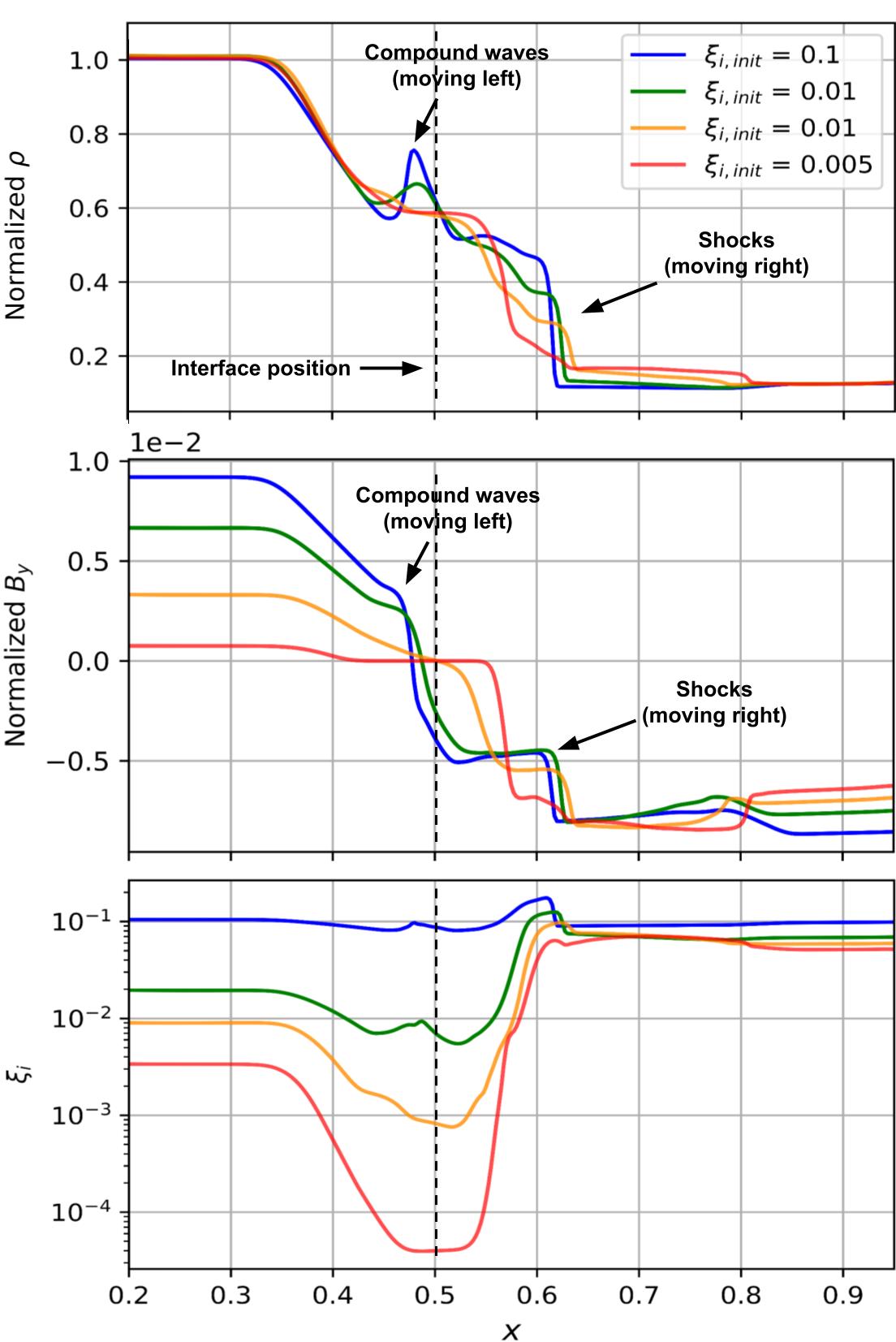}
\caption{Comparison of the \(t = 0.1\) normalized density (\textbf{top}), normalized \(B_y\) (\textbf{middle}) and ionization fraction (\textbf{bottom}) profiles of the BW test simulated with resistive effects for different initial ionization fractions. Notice the decay of electromagnetic effects with reduced ionization, showcased by the flattening of the central compound wave, which is a wave type introduced by the magnetic influences.}
\label{fig: BW resis}
\end{figure}

Finally, we examine our approach's ability to capture the enhanced resistive effects introduced by neutrals. As detailed in Appendix \ref{sec: resis map}, neutrals dramatically increase plasma resistivity as their presence reduces the number of free electrons and introduces an additional mechanism for collisions. Unlike traditional MHD simulations, our approach uses TabEoS to obtain resistivity specific to the ionization fraction and thermodynamic state of each cell, leading to a non-uniform distribution of resistivity, and further nonlinear interactions that can greatly alter the evolution of the plasma. This allows our approach to capture additional resistive phenomena inaccessible to traditional MHD simulations, such as the incitement of magnetic reconnection due to local thermodynamic fluctuations.   

We begin by again utilizing the Brio-Wu shock tube test. The relative-motion between ions and neutrals is turned off for the results in this section to isolate the effects of resistivity and allow for better comparison with literature. 

\begin{figure}[t] 
\centering    
\includegraphics[width=0.4\textwidth]{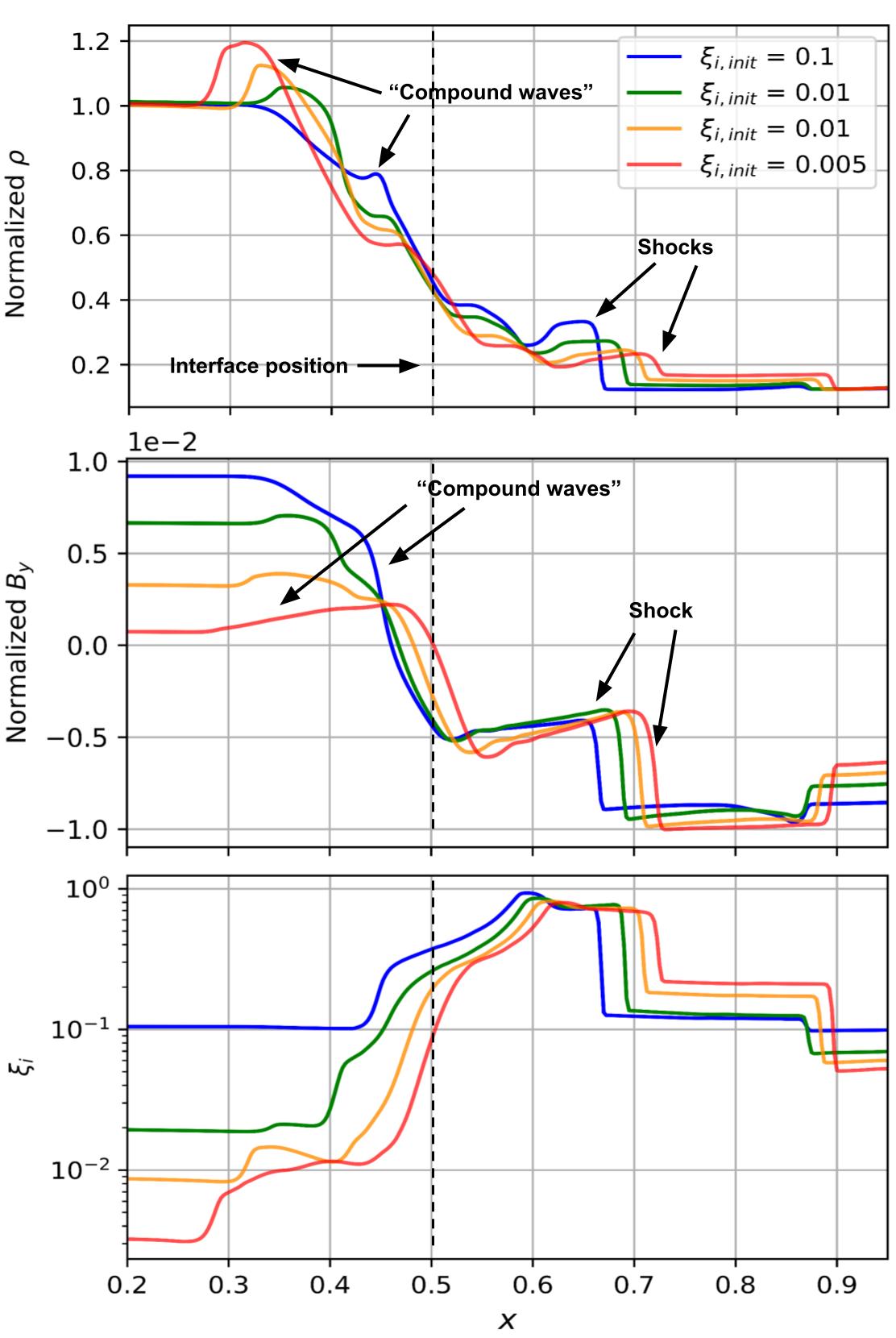}
\caption{Same comparison and profiles as Fig.~\ref{fig: BW resis} but with Joule heating. Notice the significant increase in the ionization fraction of the post-shock region, and the faster movement of the waves (shocks now at \(x \sim 0.7\), whilst at \(x \sim 0.6\) in Fig.~\ref{fig: BW resis}).}
\label{fig: BW resis JH}
\end{figure}

Fig.~\ref{fig: BW resis} and Fig.~\ref{fig: BW resis JH} shows the comparison of density, \(B_y\), and \(\xi_i\) profiles for the test initialized with progressively lower ionization fractions. The results in Fig.~\ref{fig: BW resis} are simulated without the effects of Joule heating, whereas the results in Fig.~\ref{fig: BW resis JH} are simulated with. 

For the tests without Joule heating, we see that the density profile from the \(\xi_{i, init} = 0.1\) case resembles the non-resistive results from Fig.~\ref{fig: BW gamma study}, while its magnetic profile matches the expected wave structure from the Brio-Wu test \cite{brioWu1988}. However, as the ionization fraction is reduced, we observe the resistive dissipation of the magnetic wave structure. This is most exemplified by the flattening of the compound wave (central peak in density profile), which is exclusive to magnetized fluids \cite{brioWu1988}. The results also highlight the localized nature of resistivity in our simulations, since even within the same profile the magnetic wave structure is much more diffused in the regions with lower ionization (e.g. \(x = 0.35\) - 0.55) than in the regions where it is higher (\(x > 0.6\)).

The introduction of Joule heating, which scales with \(\eta_n\) and \(J^2\) provides the PIP with a non-linear mechanism for "self-regulating" its resistivity. In a poorly ionized plasma, the resistivity is high, leading to high levels of Joule heating. This heating increases the plasma's ionization fraction, hence increasing its ability to carry current, but also reducing its resistivity. The final balance reached is dependent on the nonlinear interaction between the \(\eta_n\), \(J^2\), and any other effects on the plasma's thermodynamic state. We see these effects captured in Fig.~\ref{fig: BW resis JH}, where the ionization fraction in the post-shock region (\(x \sim 0.55\)-0.7) is two orders of magnitude greater than those in the post-shock region of Fig.~\ref{fig: BW resis} (\(x \sim 0.5\)-0.62), and the magnetic wave structure in this region is also much less diffused. In this case, the compressional heating from the shock wave has served as an additional "push" to initiate this non-linear feedback loop. As a check of consistency, the leftmost part of the domain, where no waves have yet reached (\(x \sim 0.2\)-0.3), remains at the same ionization fraction in both figures and features the same level of dissipation in its \(B_y\) profile. 

The energy introduced by Joule heating also increases the plasma wave speeds. The shock waves in Fig.~\ref{fig: BW resis JH} are now located at \(x \sim 0.7\), while they were located at \(x \sim 0.6\) in Fig.~\ref{fig: BW resis}. They are also shifted more towards the boundary with decreasing \(\xi_{i,init}\), which, as mentioned, increases \(\eta_n\) and hence Joule heating. Similarly, the discontinuity which becomes the compound wave in an ideal MHD limit is also shifted towards the left boundary.  These trends qualitatively agree with the results presented by Navarro et al. simulated using a resistive MHD simulation with constant resistivity \cite{navarro2017magnus}.

\begin{figure*}[hbtp] 
\centering    
\includegraphics[width=0.9\textwidth]{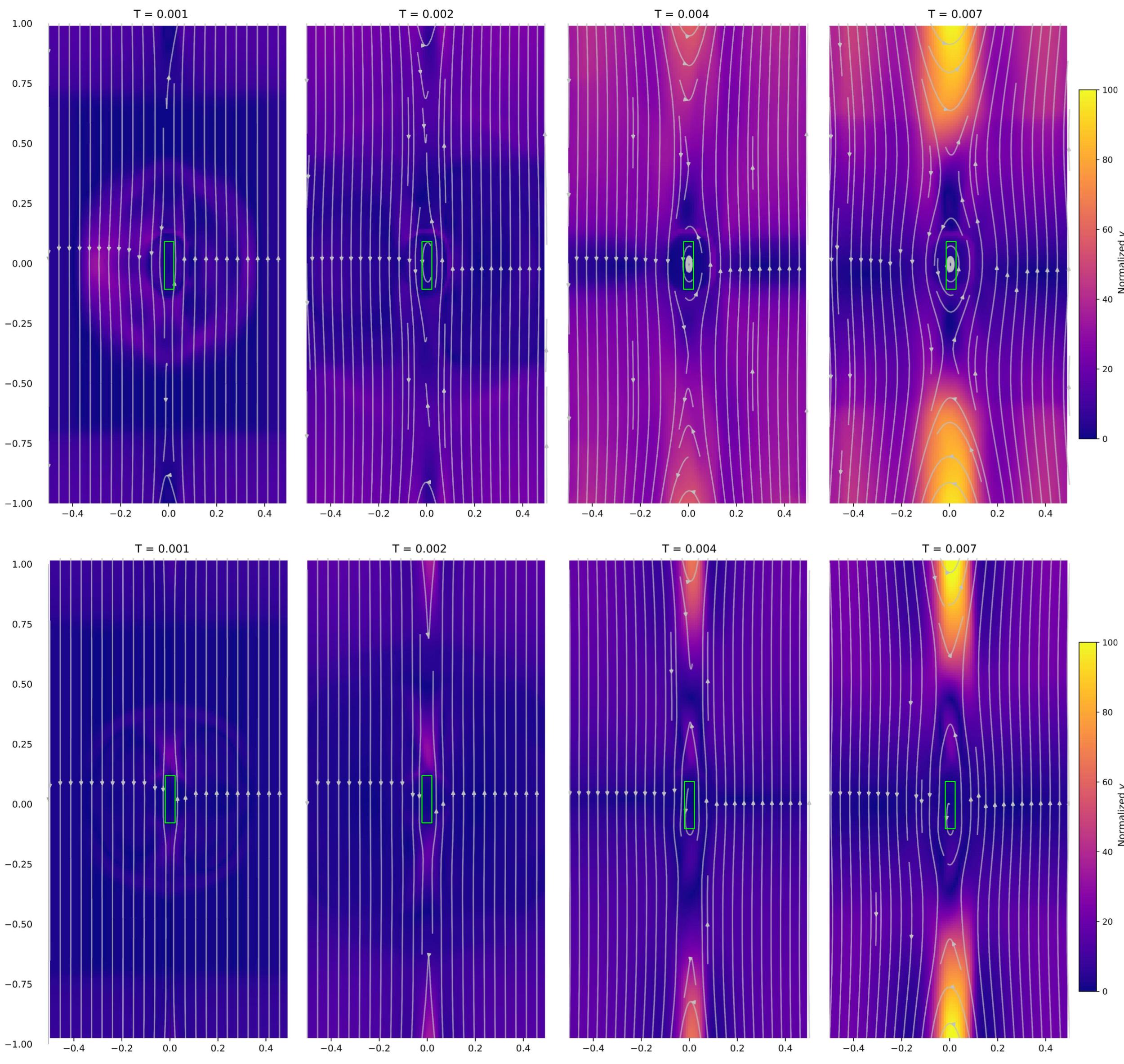}
\caption{Velocity heatmaps and magnetic field lines for the magnetic reconnection test case. The green outline marks the central region where density is set to \(2 \times 10^{-7} \ \text{kgm}^{-3}\), while it is set to \(3 \times 10^{-8} \ \text{kgm}^{-3}\) elsewhere.   
\textbf{Top}: simulated using TabEoS without explicitly adjusting resistivity. \textbf{Bottom}: simulated using resistive MHD with an ideal gas EoS. The resistivity was artificially set to match the initial profile from the TabEoS simulation, and is constant in time. Notice the formation of a tearing mode instability, and the release of acceleration of the plasma in both cases, but the significantly increased growth of the magnetic island in \(x-\)direction for the TabEoS case due to the diffusion of the central neutral region.}
\label{fig: mag recon}
\end{figure*}

We also present a demonstration of our approach's ability to capture additional neutral-induced physics in the form of a tearing-mode instability caused by a localized region of high density \cite{parra2018resistive, banerjee2024neutral_tearing_mode}. We initialize a thin current sheet as follows:
\begin{align*}
    \mathbf{v} &= \mathbf{0}, \quad p = 0.1, \quad B_x = 0, \quad \forall \quad x, y \nonumber \\
    B_y &= 
    \begin{cases}
        -1 & \text{for } x \leq -L_r \\
        \sin\left(\frac{\pi x}{2L_r} \right) & \text{if } |x| \leq L_r \\
         1 & \text{for } x > L_r
    \end{cases} \nonumber \\
    B_z &= 
    \begin{cases}
        0 & \text{for } x \leq -L_r \\
        \cos\left( \frac{\pi x}{2 L_r} \right) & \text{if } |x| \leq L_r \\
        0 & \text{for } x > L_r
    \end{cases} \nonumber
\end{align*}
where the domain of the test is \(x \in [-0.5, 0.5], \: y \in [-1,1]\), with transmissive boundary conditions. \(L_r\) defines the width of the current sheet. A central rectangular reconnection region is defined by \(x \in [-Lr, Lr], \: y \in [-2Lr,Lr]\).

Most magnetic reconnection simulations are initiated by artificially setting the resistivity to a finite constant in the central region. Using our approach, we can instead increase the density in the central region to simulate the effect of a localized high density area, which can be the result of turbulence driven perturbation, or due to the puffing of neutral particles in fusion devices. The increase in neutral density significantly increases the resistivity in the central region, inducing reconnection and leading to a tearing mode instability.

Fig.~\ref{fig: mag recon} shows the comparison of results of such a simulation initiated with \(\rho = 2 \times 10^{-7} \ \text{kgm}^{-3}\) in the central region (green box) and \(\rho = 3 \times 10^{-8}\ \text{kgm}^{-3}\) elsewhere. The results in the top row were simulated using our approach, where \(\eta_n\) is purely obtained based on the PIP's species composition and thermodynamics using TabEoS, while those from the bottom are from a traditional resistive MHD simulation using an ideal gas EoS, where the \(\eta_n\) profile is artificially set on initialization to match the resistivity profile of the first simulation and remains constant afterwards. Joule heating is turned on for these simulations. 

We see that the same general phenomena is showcased by both simulations - the high density from the central region generates a velocity wave which travels through the domain. Meanwhile, the high resistivity in the central region leads to the reconnection of magnetic field lines across the current sheet. This forms a magnetic island which grows due to the energy introduced by the reconnection, creating a tearing mode instability. Furthermore, the conversion of stored magnetic energy into thermal energy also accelerates the plasma, releasing high velocity jets in the \(y-\)direction.  

However, the growth of the instability and the energy distribution in the domain differs significantly between the two simulations. The shape of the magnetic island in the traditional resistive MHD results closely resembles the narrow rectangular shape of the central region, while in the TabEoS case, significantly more growth is observed in the \(x-\)direction, producing a more circular island. This occurs due to the diffusion of density from the central region, which has no impact on resistivity in the traditional MHD case, but leads increases the resistivity of the surrounding cells when using TabEoS, leading to an increased island growth rate. Similarly, the response of surrounding cells also leads to a greater wider distribution of velocity in the TabEoS simulation, whilst in the traditional MHD case it is again much more localized to the current sheet. 

These results serves as a rudimentary demonstration of our approach's ability to capture spontaneous resistive effects such as the formation of magnetic islands in a PIP due to localized thermodynamic perturbations. This was unavailable to usual MHD simulations which usually relies on a prescribed resistivity profile which is constant in time, or even in more advanced cases, relies on the use of Spitzer resistivity which does not account for the effect of neutrals.  

\section{Conclusion} \label{sec: conclusion}
In this paper we have addressed the need for an efficient reduced model of partially-ionized plasmas (PIPs) by presenting a single-fluid approach which captures the non-ideal effects introduced by neutrals, while remaining close to the efficiency of MHD simulations.

Central to our approach is the simulation of a single-fluid model of PIPs that captures the relative-motion between the charged and neutral particles and its effect on the bulk fluid. The efficient and accurate simulation of this system was achieved by rewriting the density, momentum, energy and induction equations in conservative form, and treating the evolution of the non-conserved relative-motion, along with any diffusive effects as separate source terms through operator splitting. The main conservative system was solved using 2nd-order finite-volume methods to increase their shock capturing abilities. The diffusive source terms were solved implicitly by using the Crank-Nicolson scheme, allowing for unconditional numerical stability and circumventing the time-step constraint of parabolic PDEs. Finally, the relative-motion is evolved using the second-order explicit Runge-Kutta scheme. 

The microscopic physics of the constituent particles of the PIP have been encapsulated in the material properties of the bulk fluid, which can change rapidly with its thermodynamic state. Hence, also critical to our approach is the use of a tabulated equation of state (TabEoS) which provides the species composition of the PIP, as well as its material properties, specific to the thermodynamic state of each cell. TabEoS is constructed using empirical data prior to the start of a simulation, and hence has minimal performance impact. For the results in this paper, it was constructed using the \( \text{Mutation}^{++} \) library, which samples from the NASA Glenn-coefficient database. However, it can also be constructed/modified using more advanced databases for specific purposes. An example is to use measurements of turbulence-modified transport coefficients to create an accurate reduced model of transport in the scrape-off layer of tokamaks.

The approach was then verified through a suite of shock-tube tests. It was shown to be capable of capturing the non-linear relationship between the degree of ionization of the PIP and its compressibility and resistivity. Furthermore, it is able to capture the relative-motion between the neutrals and the ions, with velocity profiles that matches the result of two-fluid simulations. We also demonstrate the potency of the approach through the simulation of magnetic reconnection and the growth of tearing mode instabilities induces by a local density perturbation. 

As this paper is only an initial introduction of our approach, many areas of future extensions and improvements exist. The approach can be extended to capture additional non-ideal effects. The diffusive transport of thermal conduction and viscosity have already been mentioned and can be easily added as their treatment is numerically similar to resistivity. Another crucial effect introduced by neutrals but not currently captured is the loss of energy from the plasma through radiation.  

Numerically, a detailed analysis of the eigenvalues of the simulated system would be highly valuable, as it will provide insight into the modifications to the MHD wave structure introduced by the neutral dynamics. Furthermore, efforts on the parallelization of the simulation approach would also be required to further increase its efficiency of large-domain, high-resolution problems.

Finally, the approach provided so far is generally applicable to all partially-ionized phenomena and serves as a foundation for specialization. Using the example of fusion applications, since our system automatically reduces to extended-MHD in the fully-ionized limit, it can be easily integrated with existing MHD codes such as JOREK and M3D-C1 as a module for the edge plasma. The authors of the paper are also currently developing a standalone scrape-off layer simulation based on the approach. The comparison with conventional approaches in simulating the edge of fusion plasmas will be left as a topic for future investigations.

\appendix

\section{Mapping of the Partially Ionized Regime}\label{sec: PIR mapping}

Using TabEoS, we have constructed heat maps which shows the variation in the material properties of a PIP in density-temperature space. The properties presented are either directly used in the calculation of the non-ideal effects introduced by partial-ionization or indicative of their strength, hence the heat maps are predicative of the non-ideal effects expected for a specific partially-ionized phenomenon. 

We have sampled a number density range between \(10^{17}\) to \(10^{24}\) \(\text{m}^{-3}\), which captures a wide range of environments relevant to partially-ionized phenomena, including: the edge of a tokamak reactor (\(n \sim10^{19}\)), layers of the Earth's atmosphere relevant for commercial aircraft (\(n \sim 10^{24}\)) and astronautical vehicle reentry (\(n \sim 10^{19} - 10^{23} \)) and layers of the solar atmosphere (\(n \sim 10^{17} - 10^{23} \)).

The temperature range sampled is between 0.17 to 1.72 eV (2000 to 20000 K), based around the hydrogen ionization temperature of \(\sim 0.86 \) eV (10000 K). Across this temperature range, the ionization fraction ranges from less than \(10^{-10}\) to greater than \(0.999\).

\subsection{Specific heat ratio, \(\gamma\)} \label{sec: gamma map}

\begin{figure}[h] 
\centering    
\includegraphics[width=0.5\textwidth]{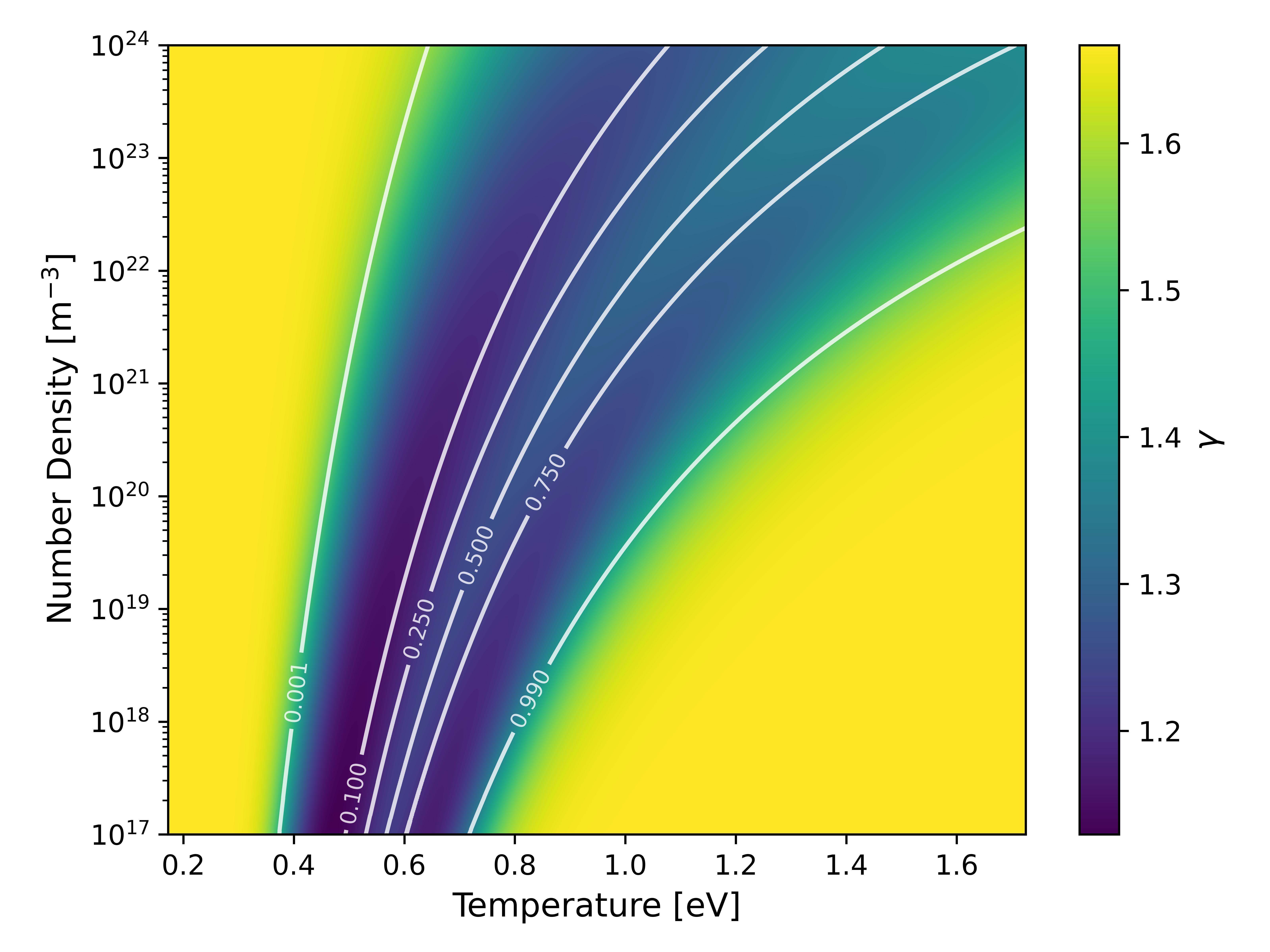}
\caption{Heatmap showing the variation of the specific heat ratio, \(\gamma\), of a PIP with number density and temperature. The white contours shows the mass fraction of ions, \(\xi_i\), in the plasma.}
\label{fig: gamma heatmap}
\end{figure}

The specific heat ratio \(\gamma\) is defined as: 
\begin{equation}
    \gamma = \frac{c_p}{c_v},
\end{equation}
where \( c_p \) and \(c_v\) are the specific heat capacities at constant pressure and volume. For ideal gases, \(\gamma\) relates the pressure of the gas and its internal energy:
\begin{equation}
    p = (\gamma - 1)\rho \varepsilon,
\end{equation}
and is also used in the calculation of its sound speed, \(c_s\):
\begin{equation}
    c_s = \sqrt{\frac{\gamma p}{\rho}}.
\end{equation}
In essence, \(\gamma\) quantifies the compressibility of a gas, or how responsive its internal energy and density are to external pressure. Hence, the use of an appropriate \(\gamma\) is essential to the accurate simulation of the gases' dynamics and temperature distribution.

For non-ideal gases, the form of the expressions above are often modified by the additional of constants offsets to account for the effects of inter-particle forces \cite{zheng2024MGEoS}, however the variation in \(\gamma\) still serves as a good indicator of the changes in the gases' compressibility.   

Equipartition theory predicts that \(\gamma\) relates to the accessible internal degree of freedom (\(f\)) of the gas as:
\begin{equation}
    \gamma = 1 + \frac{2}{f}.
\end{equation}
This is where the value of \(\gamma = \nicefrac{5}{3}\) commonly used in ideal MHD simulations originates a monoatomic ideal gas only has access to \(f = 3\). 
By accessing additional \(f\), the gas particles gain access to additional types of internal motion. These motions present new energy sinks, hence reducing the amount of fluid energy that is converted to pressure. 

In a PIP, the energy required for the ionization energy of neutrals introduces a major energy sink \cite{UCAstroNotes}. Hence \(\gamma\) will now depend on the degree of ionization of the plasma, with a more mixed plasma leading to a lower \(\gamma\), and greater compressibility.

Fig.~\ref{fig: gamma heatmap} shows the values of \(\gamma\) mapped the partially-ionized regime. We see that as expected the variation of \(\gamma\) very closely follows the variations in the ionization fraction. \(\gamma\) reaches the monoatomic equipartition prediction of \(\nicefrac{5}{3}\) at both the \(\xi_i \rightarrow 1 \) and \(\xi_i \rightarrow 0\) limits. This is because there are minimal ionization events taking place in these limits, and the plasma resembles a monoatomic ideal gas chemically (our model does not current include diatomic particles). However, \(\gamma\) changes rapidly with \(\xi_i\) between the limits, reaching a minima of \(\sim 1.15\). Hence, significant modifications of the plasma compressibility are expected in this region, necessitating the use of a more advanced EoS.

\subsection{Collisionality} \label{sec: collisionality study}

\begin{figure*}[] 
\centering    
\includegraphics[width=0.75\textwidth]{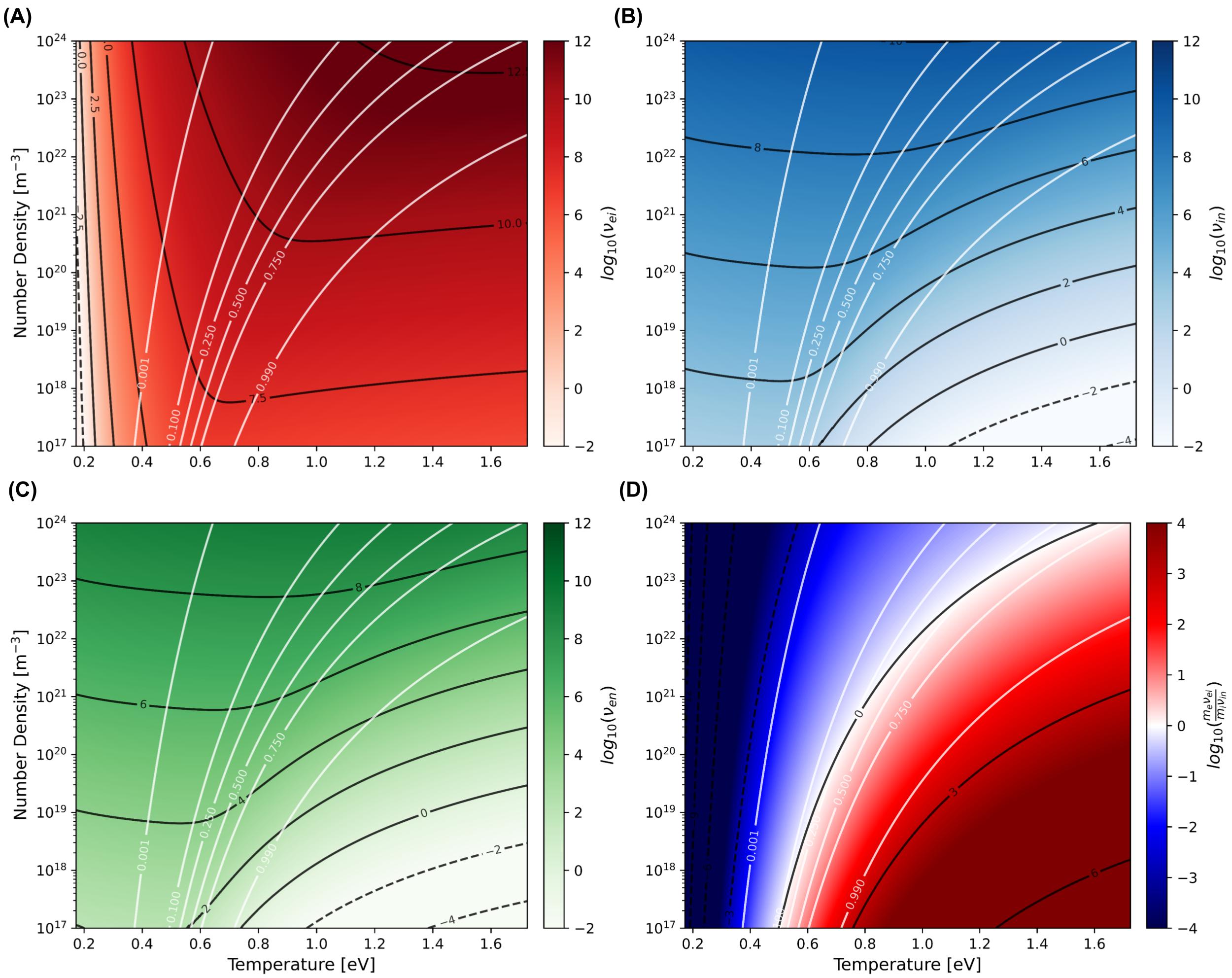}
\caption{Heat maps showing the variation of inter-species collision frequencies, \(\nu_{ab}\), of a PIP with number density and temperature. The black contours are for \(\nu\), while the white contours shows the ionization fraction, \(\xi_i\), of the plasma. \textbf{(A)}: ion-electron collision frequency, \(\nu_{ei}\), \textbf{(B)}: ion-neutral collision frequency, \(\nu_{in}\), \textbf{(C)}: electron-neutral collision frequency, \(\nu_{en}\), \textbf{(D)}: The ratio \( \frac{m_e\nu_{ei}}{m_i\nu_{in}}\), which indicates the relative collisional coupling strength between the charged particles and between the ions and the neutrals.}
\label{fig: collision freq heatmaps}

\includegraphics[width=0.75\textwidth]{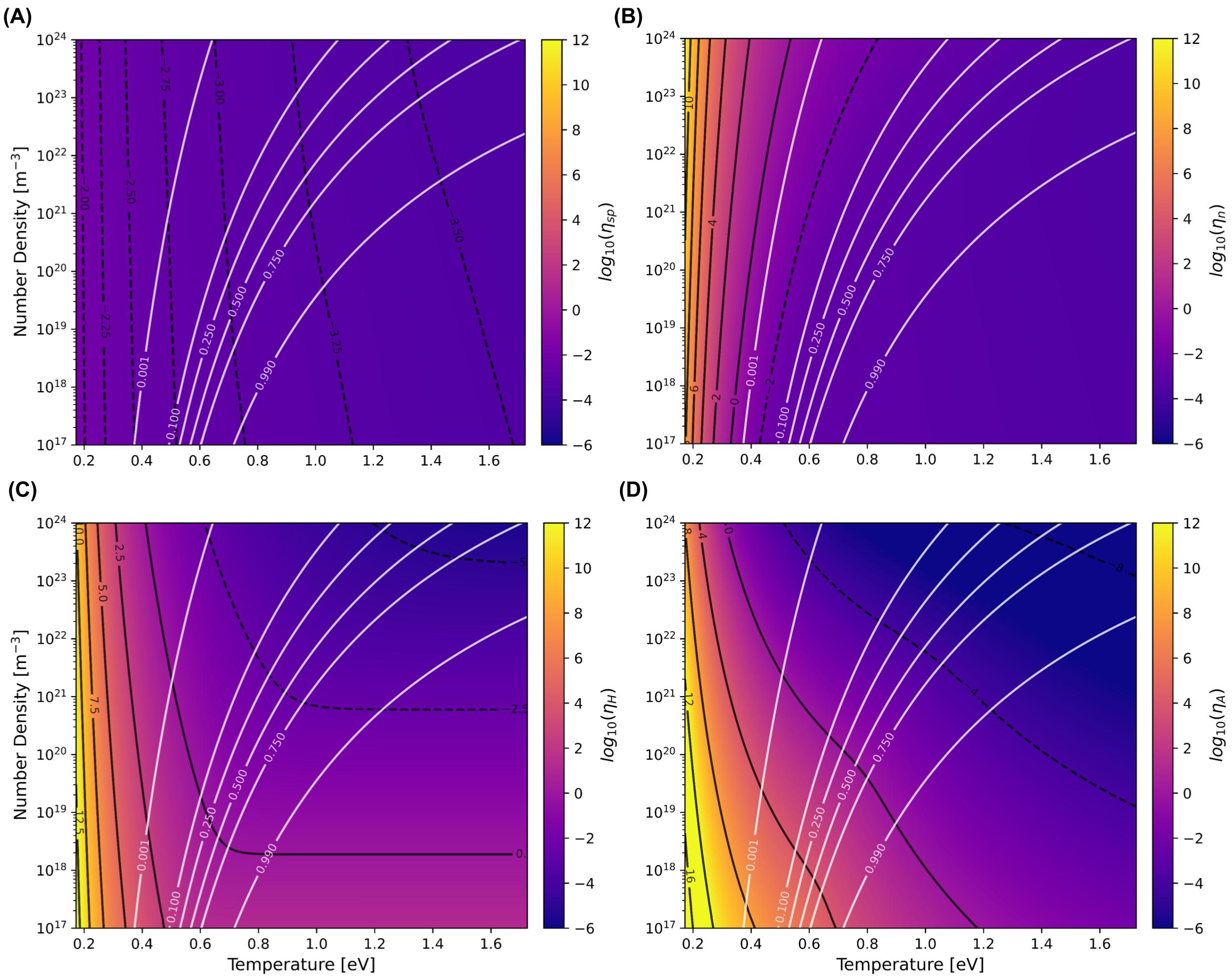}
\caption{Heat maps showing the variation of resistivites of a PIP with number density and temperature. The black contours are for \(\mathcal{\eta}\), while the white contours shows the ionization fraction, \(\xi_i\), of the plasma. \textbf{(A)}: Spitzer resistivity, \(\eta_{sp}\), \textbf{(B)}: Ohmic resistivity accounting for neutral collisions, \(\eta_n\), \textbf{(C)}: Hall resistivity, \(\eta_H\), \textbf{(D)}: Ambipolar resistivity, \(\eta_A\). }
\label{fig: resis heatmaps}

\end{figure*}

Fig.~\ref{fig: collision freq heatmaps} shows how the inter-species collision frequencies vary across the PIR. The charged collision frequency have been calculated as
\begin{equation}
    \nu_{ei} = \frac{e^4 n_i \text{ln}\Lambda}{3 \epsilon_0^2 m_e^2} \left( \frac{m_e}{2 \pi k_B T} \right)^{\nicefrac{3}{2}}
\end{equation}
where \(\text{ln}\Lambda\) is the Coulomb Logarithm \cite{krommes2019CLog}. The neutral collision frequencies are calculated as:
\begin{gather}
    \nu_{in} = n_n \sqrt{\frac{8k_BT}{\pi m_{in}}} \Sigma_{in} \\
    \nu_{en} = n_n \sqrt{\frac{8k_BT}{\pi m_{in}}} \Sigma_{en}
\end{gather}
where \(m_{ab}\) is the reduced mass of particles \(a\) and \(b\) and \(\Sigma_{ab}\) is their collisional cross section\cite{spitzer1956physics, vranjes2013collisions}, found to be \(\Sigma_{in} = 5 \times 10^{-19} \; \text{m}^2\) and \(\Sigma_{en} = 10^{-19} \; \text{m}^2\).

As expected, for the same temperature, \(\nu\) increases with the total number density of the plasma for all three inter-species combinations. 

The temperature variation is more nuanced as the number density of the individual species is also a function of \(T\). Fig.~\ref{fig: collision freq heatmaps}A shows that for \(\nu_{ei}\), the \(n_i\) dependence strongly dominants at low \(T\) and low \(\xi_i\), leading to an increase in collisionality with \(T\). This dependence wanes as the plasma becomes more ionized, reaching a limit at \(\xi_i \sim 0.7\), after which we observe a dramatic shift in the contour of the heat map, indicating the start of the \(\frac{1}{T^{\nicefrac{3}{2}}} \) dominance, and the usual inverse relationship between collisionality and \(T\).

For the neutral collisions, Fig.~\ref{fig: collision freq heatmaps}B-C show that at low \(T\), where there is an abundance of neutral particles in the plasma, \(\nu_{in}\) and \(\nu_{en}\) followed the \(T^{\nicefrac{1}{2}}\) dependence. However, as the fraction of neutral particles fell below 50\%, the \(n_n\) dependency takes over leading to the decrease in \(\nu_{in}\) and \(\nu_{en}\) with \(T\).

The final heat map shows the ratio between \( \frac{\nu_{ei}m_e}{m_i}\) and \(\nu_{in}\), which quantifies the coupling strength between the charged particles, and between the ions and the neutrals. The additional factor of \(\frac{m_e}{m_i}\) originates from the kinematic mechanism where the amount of energy transferred in a collision scales with the mass ratio of the particles \cite{braginskii1965}. 

We see that the ratio is highly correlated with \(\xi_i\). The heat map shows that for a PIP ranging between \(0.01 < \xi_i < 0.9\), the ions actually experience similar levels of coupling with the neutrals and with the electrons. This makes it difficult to justify the use of a two-fluid simulation of PIPs in this range, as either a full three-fluid model should be used for phenomena of time scale similar (or smaller) than \(\frac{1}{\nu_{in}}\), or a single-fluid model should be when the phenomenon time scale \(\gg \frac{1}{\nu_{in}}\) in the interest of efficiency.

\subsection{Resistivity} \label{sec: resis map}

Fig.~\ref{fig: resis heatmaps} shows the variation of different types of resistivities for a PIP. We have included Ohmic resistivity modified by neutrals, \(\eta_n\), Hall resistivity \cite{pandey2008hall}, \(\eta_{H}\), and Ambipolar resistivity \cite{ramshaw1991ambipolar}, \(\eta_{A}\), which contribute the magnetic induction equation as:
\begin{equation}
    \frac{\partial \mathbf{B}}{\partial t} = \nabla \times \left[ \mathbf{v} \times \mathbf{B} - \eta_n \mathbf{J} - \frac{\eta_H}{|B|} \mathbf{J} \times \mathbf{B} + \frac{\eta_A}{B^2} (\mathbf{J}\times \mathbf{B})\times \mathbf{B}\right].
\end{equation}
The commonly used Spitzer resistivity, \(\eta_{sp}\), was also included for reference. The resistivities are calculated as follows \cite{khomenko2014fluid}:
\begin{align}
    \eta_n &= \frac{\alpha_{ei} + \alpha_{en}}{e^2 n_e^2}   \\
    \eta_H &= \frac{B}{en_e} \\
    \eta_A &= \frac{\xi_n^2}{\alpha_{en} + \alpha_{in}} B^2.
\end{align}

We note the \(\mathbf{J} \times \mathbf{B}\) dependence of the Hall and Ambipolar terms. As such, an assumption of the current-perpendicular magnetic field strength is required for their calculation. We have assumed \(B = 0.3 \: \text{T}\), the polodial field strength of a standard modern tokamak. 

We see that the inclusion of neutral collision leads to a dramatic increase in \(\eta_n\) compared to \(\eta_{sp}\) at low ionization. For \(T = 0.2\)eV, \(\eta_n\) is between 8 to 10 orders of magnitude greater than \(\eta_{sp}\).

Unlike Ohmic resistivity, \(\eta_H\) and \(\eta_A\) scales inversely with number density. We see that at low number density and low temperature, these effects are up to 10 orders of magnitude greater than \(\eta_n\). \(\eta_A\) drops off rapidly as temperature and density increase, while the \(\eta_H\) remains comparable to \(\eta_n\). Hence, the inclusion of these terms in future simulations are crucial for the accurate simulation of magnetic diffusion in PIPs.

\subsection{Timescale comparisons} \label{sec: time scale comparisons}

The analysis described above can also be combined with the characteristics time and length scales of a specific partially-ionized phenomena to reveal the dominant physics expected. Let us consider the example of Alfvénic timescale within tokamak reactors, which characterizes the growth rate of MHD instabilities. The Alfvén time \(\tau_A\) is given as
\begin{equation}
    \tau_A = \frac{L}{v_A}, \quad v_A = \frac{B}{\sqrt{u_0 \rho}}
\end{equation}
where \(v_A\) is the Alfvén speed, and \(\mu_0\) is the permeability of free space. The length scale \(L\) of the phenomena is taken as the size of the fusion reactor. In particular, we will assume a configuration akin to the International Thermonuclear Experimental Reactor (ITER) , which has a magnetic field strength of \(B \approx 5\)T, and a minor radius of \(L \approx 3\)m \cite{aymar2002iter}.

\begin{figure}[h] 
\centering    
\includegraphics[width=0.5\textwidth]{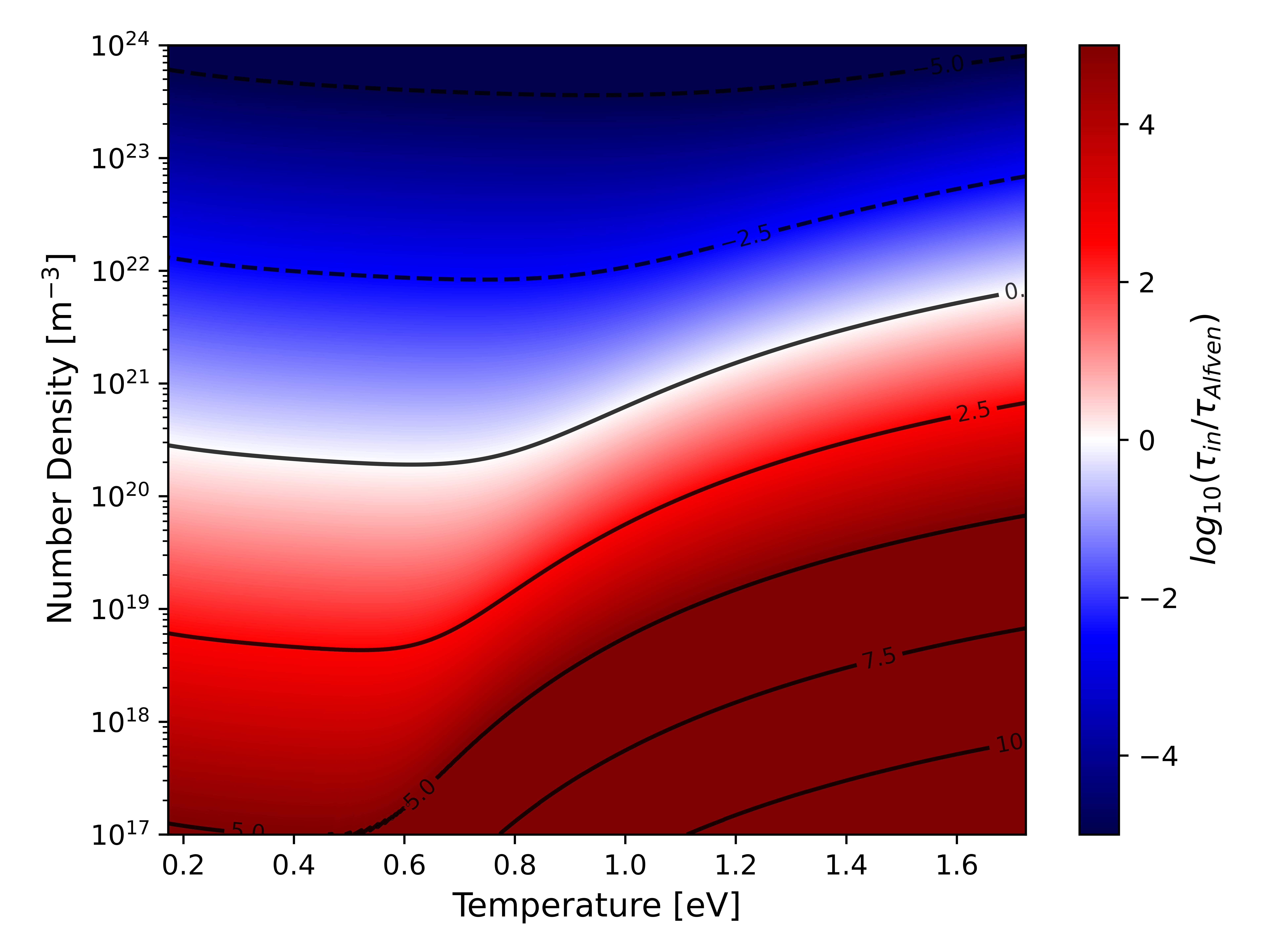}
\caption{Heatmap showing variations in the ratio between the ion-neutral collision time, \(\tau_{in}\), and the Alfvén time scale \(\tau_{A}\), calculated with an ITER-like configuration of \(B\) = 5T and length scale \(L\) = 5m, with different thermodynamic states across the PIR. }
\label{fig: time scale comparison}
\end{figure}

Fig.~\ref{fig: time scale comparison} shows the comparisons between \(\tau_A\) and \(\tau_{in}\), which establishes whether the neutrals are coupled to the electromagnetically driven dynamics of the charged species. We see that in ITER-like conditions, a timescale of \(\tau_A\) is insufficient for the collisional coupling of neutrals with the ions if \(n \leq 10^{20} \). Both the tokamak core (\(n \approx 10^{20}\)) and edge (\(n \approx 10^{19}\)) are below this limit. Consequently, a simulation needs to account for the relative-motion between the ions and the neutrals to accurately simulate the growth of MHD instabilities involving neutrals, justifying the need for our single-fluid model over traditional extended-MHD models.

\nocite{*}
\bibliography{references}% Produces the bibliography via BibTeX.

\begin{thebibliography}{79}%
\makeatletter
\providecommand \@ifxundefined [1]{%
 \@ifx{#1\undefined}
}%
\providecommand \@ifnum [1]{%
 \ifnum #1\expandafter \@firstoftwo
 \else \expandafter \@secondoftwo
 \fi
}%
\providecommand \@ifx [1]{%
 \ifx #1\expandafter \@firstoftwo
 \else \expandafter \@secondoftwo
 \fi
}%
\providecommand \natexlab [1]{#1}%
\providecommand \enquote  [1]{``#1''}%
\providecommand \bibnamefont  [1]{#1}%
\providecommand \bibfnamefont [1]{#1}%
\providecommand \citenamefont [1]{#1}%
\providecommand \href@noop [0]{\@secondoftwo}%
\providecommand \href [0]{\begingroup \@sanitize@url \@href}%
\providecommand \@href[1]{\@@startlink{#1}\@@href}%
\providecommand \@@href[1]{\endgroup#1\@@endlink}%
\providecommand \@sanitize@url [0]{\catcode `\\12\catcode `\$12\catcode `\&12\catcode `\#12\catcode `\^12\catcode `\_12\catcode `\%12\relax}%
\providecommand \@@startlink[1]{}%
\providecommand \@@endlink[0]{}%
\providecommand \url  [0]{\begingroup\@sanitize@url \@url }%
\providecommand \@url [1]{\endgroup\@href {#1}{\urlprefix }}%
\providecommand \urlprefix  [0]{URL }%
\providecommand \Eprint [0]{\href }%
\providecommand \doibase [0]{http://dx.doi.org/}%
\providecommand \selectlanguage [0]{\@gobble}%
\providecommand \bibinfo  [0]{\@secondoftwo}%
\providecommand \bibfield  [0]{\@secondoftwo}%
\providecommand \translation [1]{[#1]}%
\providecommand \BibitemOpen [0]{}%
\providecommand \bibitemStop [0]{}%
\providecommand \bibitemNoStop [0]{.\EOS\space}%
\providecommand \EOS [0]{\spacefactor3000\relax}%
\providecommand \BibitemShut  [1]{\csname bibitem#1\endcsname}%
\let\auto@bib@innerbib\@empty
%</preamble>
\bibitem [{\citenamefont {Lazerson}\ \emph {et~al.}(2016)\citenamefont {Lazerson}, \citenamefont {Loizu}, \citenamefont {Hirshman},\ and\ \citenamefont {Hudson}}]{lazerson2016VMEC}%
  \BibitemOpen
  \bibfield  {author} {\bibinfo {author} {\bibfnamefont {S.~A.}\ \bibnamefont {Lazerson}}, \bibinfo {author} {\bibfnamefont {J.}~\bibnamefont {Loizu}}, \bibinfo {author} {\bibfnamefont {S.}~\bibnamefont {Hirshman}}, \ and\ \bibinfo {author} {\bibfnamefont {S.~R.}\ \bibnamefont {Hudson}},\ }\bibfield  {title} {\enquote {\bibinfo {title} {Verification of the ideal magnetohydrodynamic response at rational surfaces in the vmec code},}\ }\href@noop {} {\bibfield  {journal} {\bibinfo  {journal} {Physics of Plasmas}\ }\textbf {\bibinfo {volume} {23}} (\bibinfo {year} {2016})}\BibitemShut {NoStop}%
\bibitem [{\citenamefont {Anninos}, \citenamefont {Fragile},\ and\ \citenamefont {Salmonson}(2005)}]{anninos2005cosmos++}%
  \BibitemOpen
  \bibfield  {author} {\bibinfo {author} {\bibfnamefont {P.}~\bibnamefont {Anninos}}, \bibinfo {author} {\bibfnamefont {P.~C.}\ \bibnamefont {Fragile}}, \ and\ \bibinfo {author} {\bibfnamefont {J.~D.}\ \bibnamefont {Salmonson}},\ }\bibfield  {title} {\enquote {\bibinfo {title} {Cosmos++: relativistic magnetohydrodynamics on unstructured grids with local adaptive refinement},}\ }\href@noop {} {\bibfield  {journal} {\bibinfo  {journal} {The Astrophysical Journal}\ }\textbf {\bibinfo {volume} {635}},\ \bibinfo {pages} {723} (\bibinfo {year} {2005})}\BibitemShut {NoStop}%
\bibitem [{\citenamefont {Gonz{\'a}lez-Avil{\'e}s}\ \emph {et~al.}(2015)\citenamefont {Gonz{\'a}lez-Avil{\'e}s}, \citenamefont {Cruz-Osorio}, \citenamefont {Lora-Clavijo},\ and\ \citenamefont {Guzm{\'a}n}}]{gonzalez2015solarIdealMHD}%
  \BibitemOpen
  \bibfield  {author} {\bibinfo {author} {\bibfnamefont {J.}~\bibnamefont {Gonz{\'a}lez-Avil{\'e}s}}, \bibinfo {author} {\bibfnamefont {A.}~\bibnamefont {Cruz-Osorio}}, \bibinfo {author} {\bibfnamefont {F.}~\bibnamefont {Lora-Clavijo}}, \ and\ \bibinfo {author} {\bibfnamefont {F.}~\bibnamefont {Guzm{\'a}n}},\ }\bibfield  {title} {\enquote {\bibinfo {title} {Newtonian cafe: a new ideal mhd code to study the solar atmosphere},}\ }\href@noop {} {\bibfield  {journal} {\bibinfo  {journal} {Monthly Notices of the Royal Astronomical Society}\ }\textbf {\bibinfo {volume} {454}},\ \bibinfo {pages} {1871--1885} (\bibinfo {year} {2015})}\BibitemShut {NoStop}%
\bibitem [{\citenamefont {Fusheng}\ \emph {et~al.}(2018)\citenamefont {Fusheng}, \citenamefont {Xiangteng}, \citenamefont {Han},\ and\ \citenamefont {Zhang}}]{fusheng2018lightningIdeal}%
  \BibitemOpen
  \bibfield  {author} {\bibinfo {author} {\bibfnamefont {W.}~\bibnamefont {Fusheng}}, \bibinfo {author} {\bibfnamefont {M.}~\bibnamefont {Xiangteng}}, \bibinfo {author} {\bibfnamefont {C.}~\bibnamefont {Han}}, \ and\ \bibinfo {author} {\bibfnamefont {Y.}~\bibnamefont {Zhang}},\ }\bibfield  {title} {\enquote {\bibinfo {title} {Evolution simulation of lightning discharge based on a magnetohydrodynamics method},}\ }\href@noop {} {\bibfield  {journal} {\bibinfo  {journal} {Plasma Science and Technology}\ }\textbf {\bibinfo {volume} {20}},\ \bibinfo {pages} {075301} (\bibinfo {year} {2018})}\BibitemShut {NoStop}%
\bibitem [{\citenamefont {Morley}\ \emph {et~al.}(2004)\citenamefont {Morley}, \citenamefont {Smolentsev}, \citenamefont {Munipalli}, \citenamefont {Ni}, \citenamefont {Gao},\ and\ \citenamefont {Abdou}}]{morley2004HIMAG}%
  \BibitemOpen
  \bibfield  {author} {\bibinfo {author} {\bibfnamefont {N.}~\bibnamefont {Morley}}, \bibinfo {author} {\bibfnamefont {S.}~\bibnamefont {Smolentsev}}, \bibinfo {author} {\bibfnamefont {R.}~\bibnamefont {Munipalli}}, \bibinfo {author} {\bibfnamefont {M.-J.}\ \bibnamefont {Ni}}, \bibinfo {author} {\bibfnamefont {D.}~\bibnamefont {Gao}}, \ and\ \bibinfo {author} {\bibfnamefont {M.}~\bibnamefont {Abdou}},\ }\bibfield  {title} {\enquote {\bibinfo {title} {Progress on the modeling of liquid metal, free surface, mhd flows for fusion liquid walls},}\ }\href@noop {} {\bibfield  {journal} {\bibinfo  {journal} {Fusion Engineering and Design}\ }\textbf {\bibinfo {volume} {72}},\ \bibinfo {pages} {3--34} (\bibinfo {year} {2004})}\BibitemShut {NoStop}%
\bibitem [{\citenamefont {Greenwald}, \citenamefont {Shumlak},\ and\ \citenamefont {Anderson}(2024)}]{greenwald2024preface}%
  \BibitemOpen
  \bibfield  {author} {\bibinfo {author} {\bibfnamefont {M.}~\bibnamefont {Greenwald}}, \bibinfo {author} {\bibfnamefont {U.}~\bibnamefont {Shumlak}}, \ and\ \bibinfo {author} {\bibfnamefont {D.}~\bibnamefont {Anderson}},\ }\bibfield  {title} {\enquote {\bibinfo {title} {Preface to special issue: Private fusion research: Opportunities and challenges in plasma science},}\ }\href@noop {} {\bibfield  {journal} {\bibinfo  {journal} {Physics of Plasmas}\ }\textbf {\bibinfo {volume} {31}},\ \bibinfo {pages} {090401} (\bibinfo {year} {2024})}\BibitemShut {NoStop}%
\bibitem [{\citenamefont {Chapman}, \citenamefont {Bestwick},\ and\ \citenamefont {Methven}(2023)}]{chapman2023public}%
  \BibitemOpen
  \bibfield  {author} {\bibinfo {author} {\bibfnamefont {I.}~\bibnamefont {Chapman}}, \bibinfo {author} {\bibfnamefont {T.}~\bibnamefont {Bestwick}}, \ and\ \bibinfo {author} {\bibfnamefont {P.}~\bibnamefont {Methven}},\ }\bibfield  {title} {\enquote {\bibinfo {title} {Public--private partnership in the uk fusion program},}\ }\href@noop {} {\bibfield  {journal} {\bibinfo  {journal} {Physics of Plasmas}\ }\textbf {\bibinfo {volume} {30}} (\bibinfo {year} {2023})}\BibitemShut {NoStop}%
\bibitem [{\citenamefont {Kingham}\ and\ \citenamefont {Gryaznevich}(2024)}]{kingham2024TE}%
  \BibitemOpen
  \bibfield  {author} {\bibinfo {author} {\bibfnamefont {D.}~\bibnamefont {Kingham}}\ and\ \bibinfo {author} {\bibfnamefont {M.}~\bibnamefont {Gryaznevich}},\ }\bibfield  {title} {\enquote {\bibinfo {title} {The spherical tokamak path to fusion power: Opportunities and challenges for development via public--private partnerships},}\ }\href@noop {} {\bibfield  {journal} {\bibinfo  {journal} {Physics of Plasmas}\ }\textbf {\bibinfo {volume} {31}} (\bibinfo {year} {2024})}\BibitemShut {NoStop}%
\bibitem [{\citenamefont {Creely}\ \emph {et~al.}(2023)\citenamefont {Creely}, \citenamefont {Brunner}, \citenamefont {Mumgaard}, \citenamefont {Reinke}, \citenamefont {Segal}, \citenamefont {Sorbom},\ and\ \citenamefont {Greenwald}}]{creely2023CFS}%
  \BibitemOpen
  \bibfield  {author} {\bibinfo {author} {\bibfnamefont {A.}~\bibnamefont {Creely}}, \bibinfo {author} {\bibfnamefont {D.}~\bibnamefont {Brunner}}, \bibinfo {author} {\bibfnamefont {R.}~\bibnamefont {Mumgaard}}, \bibinfo {author} {\bibfnamefont {M.}~\bibnamefont {Reinke}}, \bibinfo {author} {\bibfnamefont {M.}~\bibnamefont {Segal}}, \bibinfo {author} {\bibfnamefont {B.}~\bibnamefont {Sorbom}}, \ and\ \bibinfo {author} {\bibfnamefont {M.}~\bibnamefont {Greenwald}},\ }\bibfield  {title} {\enquote {\bibinfo {title} {Sparc as a platform to advance tokamak science},}\ }\href@noop {} {\bibfield  {journal} {\bibinfo  {journal} {Physics of Plasmas}\ }\textbf {\bibinfo {volume} {30}} (\bibinfo {year} {2023})}\BibitemShut {NoStop}%
\bibitem [{\citenamefont {Levitt}\ \emph {et~al.}(2023)\citenamefont {Levitt}, \citenamefont {Meier}, \citenamefont {Umstattd}, \citenamefont {Barhydt}, \citenamefont {Datta}, \citenamefont {Liekhus-Schmaltz}, \citenamefont {Sutherland},\ and\ \citenamefont {Nelson}}]{levitt2023zap}%
  \BibitemOpen
  \bibfield  {author} {\bibinfo {author} {\bibfnamefont {B.}~\bibnamefont {Levitt}}, \bibinfo {author} {\bibfnamefont {E.}~\bibnamefont {Meier}}, \bibinfo {author} {\bibfnamefont {R.}~\bibnamefont {Umstattd}}, \bibinfo {author} {\bibfnamefont {J.}~\bibnamefont {Barhydt}}, \bibinfo {author} {\bibfnamefont {I.}~\bibnamefont {Datta}}, \bibinfo {author} {\bibfnamefont {C.}~\bibnamefont {Liekhus-Schmaltz}}, \bibinfo {author} {\bibfnamefont {D.}~\bibnamefont {Sutherland}}, \ and\ \bibinfo {author} {\bibfnamefont {B.}~\bibnamefont {Nelson}},\ }\bibfield  {title} {\enquote {\bibinfo {title} {The zap energy approach to commercial fusion},}\ }\href@noop {} {\bibfield  {journal} {\bibinfo  {journal} {Physics of Plasmas}\ }\textbf {\bibinfo {volume} {30}} (\bibinfo {year} {2023})}\BibitemShut {NoStop}%
\bibitem [{\citenamefont {Miyazawa}\ and\ \citenamefont {Goto}(2023)}]{miyazawa2023Helical}%
  \BibitemOpen
  \bibfield  {author} {\bibinfo {author} {\bibfnamefont {J.}~\bibnamefont {Miyazawa}}\ and\ \bibinfo {author} {\bibfnamefont {T.}~\bibnamefont {Goto}},\ }\bibfield  {title} {\enquote {\bibinfo {title} {Development of steady-state fusion reactor by helical fusion},}\ }\href@noop {} {\bibfield  {journal} {\bibinfo  {journal} {Physics of Plasmas}\ }\textbf {\bibinfo {volume} {30}} (\bibinfo {year} {2023})}\BibitemShut {NoStop}%
\bibitem [{\citenamefont {Mehlhorn}(2024)}]{mehlhorn2024kms}%
  \BibitemOpen
  \bibfield  {author} {\bibinfo {author} {\bibfnamefont {T.~A.}\ \bibnamefont {Mehlhorn}},\ }\bibfield  {title} {\enquote {\bibinfo {title} {From kms fusion to hb11 energy and xcimer energy, a personal 50 year ife perspective},}\ }\href@noop {} {\bibfield  {journal} {\bibinfo  {journal} {Physics of Plasmas}\ }\textbf {\bibinfo {volume} {31}} (\bibinfo {year} {2024})}\BibitemShut {NoStop}%
\bibitem [{\citenamefont {Stone}\ \emph {et~al.}(2008{\natexlab{a}})\citenamefont {Stone}, \citenamefont {Lindenmoyer}, \citenamefont {French}, \citenamefont {Musk}, \citenamefont {Gump}, \citenamefont {Kathuria}, \citenamefont {Miller}, \citenamefont {Sirangelo},\ and\ \citenamefont {Pickens}}]{stone2008nasaCOTS}%
  \BibitemOpen
  \bibfield  {author} {\bibinfo {author} {\bibfnamefont {D.}~\bibnamefont {Stone}}, \bibinfo {author} {\bibfnamefont {A.}~\bibnamefont {Lindenmoyer}}, \bibinfo {author} {\bibfnamefont {G.}~\bibnamefont {French}}, \bibinfo {author} {\bibfnamefont {E.}~\bibnamefont {Musk}}, \bibinfo {author} {\bibfnamefont {D.}~\bibnamefont {Gump}}, \bibinfo {author} {\bibfnamefont {C.}~\bibnamefont {Kathuria}}, \bibinfo {author} {\bibfnamefont {C.}~\bibnamefont {Miller}}, \bibinfo {author} {\bibfnamefont {M.}~\bibnamefont {Sirangelo}}, \ and\ \bibinfo {author} {\bibfnamefont {T.}~\bibnamefont {Pickens}},\ }\bibfield  {title} {\enquote {\bibinfo {title} {Nasa's approach to commercial cargo and crew transportation},}\ }\href@noop {} {\bibfield  {journal} {\bibinfo  {journal} {Acta Astronautica}\ }\textbf {\bibinfo {volume} {63}},\ \bibinfo {pages} {192--197} (\bibinfo {year} {2008}{\natexlab{a}})}\BibitemShut {NoStop}%
\bibitem [{\citenamefont {Vozoff}\ and\ \citenamefont {Couluris}(2008)}]{vozoff2008spacex}%
  \BibitemOpen
  \bibfield  {author} {\bibinfo {author} {\bibfnamefont {M.}~\bibnamefont {Vozoff}}\ and\ \bibinfo {author} {\bibfnamefont {J.}~\bibnamefont {Couluris}},\ }\bibfield  {title} {\enquote {\bibinfo {title} {Spacex products-advancing the use of space},}\ }in\ \href@noop {} {\emph {\bibinfo {booktitle} {AIAA SPACE 2008 conference \& exposition}}}\ (\bibinfo {year} {2008})\ p.\ \bibinfo {pages} {7836}\BibitemShut {NoStop}%
\bibitem [{\citenamefont {Bieniawski}\ \emph {et~al.}(2022)\citenamefont {Bieniawski}, \citenamefont {Lewis}, \citenamefont {Friia}, \citenamefont {Mahajan}, \citenamefont {Somervill}, \citenamefont {Mamidipudi},\ and\ \citenamefont {Dakin}}]{bieniawski2022blue_origin}%
  \BibitemOpen
  \bibfield  {author} {\bibinfo {author} {\bibfnamefont {S.~R.}\ \bibnamefont {Bieniawski}}, \bibinfo {author} {\bibfnamefont {B.}~\bibnamefont {Lewis}}, \bibinfo {author} {\bibfnamefont {B.}~\bibnamefont {Friia}}, \bibinfo {author} {\bibfnamefont {A.}~\bibnamefont {Mahajan}}, \bibinfo {author} {\bibfnamefont {K.}~\bibnamefont {Somervill}}, \bibinfo {author} {\bibfnamefont {P.}~\bibnamefont {Mamidipudi}}, \ and\ \bibinfo {author} {\bibfnamefont {D.}~\bibnamefont {Dakin}},\ }\bibfield  {title} {\enquote {\bibinfo {title} {New shepard flight test results from blue origin de-orbit descent and landing tipping point},}\ }in\ \href@noop {} {\emph {\bibinfo {booktitle} {AIAA SCITECH 2022 Forum}}}\ (\bibinfo {year} {2022})\ p.\ \bibinfo {pages} {1829}\BibitemShut {NoStop}%
\bibitem [{\citenamefont {French}\ \emph {et~al.}(2022)\citenamefont {French}, \citenamefont {Mandy}, \citenamefont {Hunter}, \citenamefont {Mosleh}, \citenamefont {Sinclair}, \citenamefont {Beck}, \citenamefont {Seager}, \citenamefont {Petkowski}, \citenamefont {Carr}, \citenamefont {Grinspoon} \emph {et~al.}}]{french2022rocketLab}%
  \BibitemOpen
  \bibfield  {author} {\bibinfo {author} {\bibfnamefont {R.}~\bibnamefont {French}}, \bibinfo {author} {\bibfnamefont {C.}~\bibnamefont {Mandy}}, \bibinfo {author} {\bibfnamefont {R.}~\bibnamefont {Hunter}}, \bibinfo {author} {\bibfnamefont {E.}~\bibnamefont {Mosleh}}, \bibinfo {author} {\bibfnamefont {D.}~\bibnamefont {Sinclair}}, \bibinfo {author} {\bibfnamefont {P.}~\bibnamefont {Beck}}, \bibinfo {author} {\bibfnamefont {S.}~\bibnamefont {Seager}}, \bibinfo {author} {\bibfnamefont {J.~J.}\ \bibnamefont {Petkowski}}, \bibinfo {author} {\bibfnamefont {C.~E.}\ \bibnamefont {Carr}}, \bibinfo {author} {\bibfnamefont {D.~H.}\ \bibnamefont {Grinspoon}},  \emph {et~al.},\ }\bibfield  {title} {\enquote {\bibinfo {title} {Rocket lab mission to venus},}\ }\href@noop {} {\bibfield  {journal} {\bibinfo  {journal} {Aerospace}\ }\textbf {\bibinfo {volume} {9}},\ \bibinfo {pages} {445} (\bibinfo {year} {2022})}\BibitemShut {NoStop}%
\bibitem [{\citenamefont {Gates}\ \emph {et~al.}(2017)\citenamefont {Gates}, \citenamefont {Boozer}, \citenamefont {Brown}, \citenamefont {Breslau}, \citenamefont {Curreli}, \citenamefont {Landreman}, \citenamefont {Lazerson}, \citenamefont {Lore}, \citenamefont {Mynick}, \citenamefont {Neilson} \emph {et~al.}}]{gates2017StellaratorOp}%
  \BibitemOpen
  \bibfield  {author} {\bibinfo {author} {\bibfnamefont {D.}~\bibnamefont {Gates}}, \bibinfo {author} {\bibfnamefont {A.}~\bibnamefont {Boozer}}, \bibinfo {author} {\bibfnamefont {T.}~\bibnamefont {Brown}}, \bibinfo {author} {\bibfnamefont {J.}~\bibnamefont {Breslau}}, \bibinfo {author} {\bibfnamefont {D.}~\bibnamefont {Curreli}}, \bibinfo {author} {\bibfnamefont {M.}~\bibnamefont {Landreman}}, \bibinfo {author} {\bibfnamefont {S.}~\bibnamefont {Lazerson}}, \bibinfo {author} {\bibfnamefont {J.}~\bibnamefont {Lore}}, \bibinfo {author} {\bibfnamefont {H.}~\bibnamefont {Mynick}}, \bibinfo {author} {\bibfnamefont {G.}~\bibnamefont {Neilson}},  \emph {et~al.},\ }\bibfield  {title} {\enquote {\bibinfo {title} {Recent advances in stellarator optimization},}\ }\href@noop {} {\bibfield  {journal} {\bibinfo  {journal} {Nuclear Fusion}\ }\textbf {\bibinfo {volume} {57}},\ \bibinfo {pages} {126064} (\bibinfo {year} {2017})}\BibitemShut {NoStop}%
\bibitem [{\citenamefont {Felici}\ and\ \citenamefont {Sauter}(2012)}]{felici2012Actuator}%
  \BibitemOpen
  \bibfield  {author} {\bibinfo {author} {\bibfnamefont {F.}~\bibnamefont {Felici}}\ and\ \bibinfo {author} {\bibfnamefont {O.}~\bibnamefont {Sauter}},\ }\bibfield  {title} {\enquote {\bibinfo {title} {Non-linear model-based optimization of actuator trajectories for tokamak plasma profile control},}\ }\href@noop {} {\bibfield  {journal} {\bibinfo  {journal} {Plasma Physics and Controlled Fusion}\ }\textbf {\bibinfo {volume} {54}},\ \bibinfo {pages} {025002} (\bibinfo {year} {2012})}\BibitemShut {NoStop}%
\bibitem [{\citenamefont {Wehner}\ \emph {et~al.}(2019)\citenamefont {Wehner}, \citenamefont {Schuster}, \citenamefont {Boyer},\ and\ \citenamefont {Poli}}]{wehner2019transp}%
  \BibitemOpen
  \bibfield  {author} {\bibinfo {author} {\bibfnamefont {W.}~\bibnamefont {Wehner}}, \bibinfo {author} {\bibfnamefont {E.}~\bibnamefont {Schuster}}, \bibinfo {author} {\bibfnamefont {M.}~\bibnamefont {Boyer}}, \ and\ \bibinfo {author} {\bibfnamefont {F.}~\bibnamefont {Poli}},\ }\bibfield  {title} {\enquote {\bibinfo {title} {Transp-based optimization towards tokamak scenario development},}\ }\href@noop {} {\bibfield  {journal} {\bibinfo  {journal} {Fusion Engineering and Design}\ }\textbf {\bibinfo {volume} {146}},\ \bibinfo {pages} {547--550} (\bibinfo {year} {2019})}\BibitemShut {NoStop}%
\bibitem [{\citenamefont {Piccione}\ \emph {et~al.}(2020)\citenamefont {Piccione}, \citenamefont {Berkery}, \citenamefont {Sabbagh},\ and\ \citenamefont {Andreopoulos}}]{piccione2020physicsML}%
  \BibitemOpen
  \bibfield  {author} {\bibinfo {author} {\bibfnamefont {A.}~\bibnamefont {Piccione}}, \bibinfo {author} {\bibfnamefont {J.}~\bibnamefont {Berkery}}, \bibinfo {author} {\bibfnamefont {S.}~\bibnamefont {Sabbagh}}, \ and\ \bibinfo {author} {\bibfnamefont {Y.}~\bibnamefont {Andreopoulos}},\ }\bibfield  {title} {\enquote {\bibinfo {title} {Physics-guided machine learning approaches to predict the ideal stability properties of fusion plasmas},}\ }\href@noop {} {\bibfield  {journal} {\bibinfo  {journal} {Nuclear Fusion}\ }\textbf {\bibinfo {volume} {60}},\ \bibinfo {pages} {046033} (\bibinfo {year} {2020})}\BibitemShut {NoStop}%
\bibitem [{\citenamefont {Hoelzl}\ \emph {et~al.}(2021)\citenamefont {Hoelzl}, \citenamefont {Huijsmans}, \citenamefont {Pamela}, \citenamefont {Becoulet}, \citenamefont {Nardon}, \citenamefont {Artola}, \citenamefont {Nkonga}, \citenamefont {Atanasiu}, \citenamefont {Bandaru}, \citenamefont {Bhole} \emph {et~al.}}]{hoelzl2021jorek}%
  \BibitemOpen
  \bibfield  {author} {\bibinfo {author} {\bibfnamefont {M.}~\bibnamefont {Hoelzl}}, \bibinfo {author} {\bibfnamefont {G.~T.}\ \bibnamefont {Huijsmans}}, \bibinfo {author} {\bibfnamefont {S.}~\bibnamefont {Pamela}}, \bibinfo {author} {\bibfnamefont {M.}~\bibnamefont {Becoulet}}, \bibinfo {author} {\bibfnamefont {E.}~\bibnamefont {Nardon}}, \bibinfo {author} {\bibfnamefont {F.~J.}\ \bibnamefont {Artola}}, \bibinfo {author} {\bibfnamefont {B.}~\bibnamefont {Nkonga}}, \bibinfo {author} {\bibfnamefont {C.~V.}\ \bibnamefont {Atanasiu}}, \bibinfo {author} {\bibfnamefont {V.}~\bibnamefont {Bandaru}}, \bibinfo {author} {\bibfnamefont {A.}~\bibnamefont {Bhole}},  \emph {et~al.},\ }\bibfield  {title} {\enquote {\bibinfo {title} {The jorek non-linear extended mhd code and applications to large-scale instabilities and their control in magnetically confined fusion plasmas},}\ }\href@noop {} {\bibfield  {journal} {\bibinfo  {journal} {Nuclear Fusion}\ }\textbf {\bibinfo {volume} {61}},\ \bibinfo {pages} {065001} (\bibinfo
  {year} {2021})}\BibitemShut {NoStop}%
\bibitem [{\citenamefont {Ferraro}, \citenamefont {Jardin},\ and\ \citenamefont {Snyder}(2010)}]{ferraro2010M3DC1}%
  \BibitemOpen
  \bibfield  {author} {\bibinfo {author} {\bibfnamefont {N.}~\bibnamefont {Ferraro}}, \bibinfo {author} {\bibfnamefont {S.~C.}\ \bibnamefont {Jardin}}, \ and\ \bibinfo {author} {\bibfnamefont {P.}~\bibnamefont {Snyder}},\ }\bibfield  {title} {\enquote {\bibinfo {title} {Ideal and resistive edge stability calculations with m3d-c1},}\ }\href@noop {} {\bibfield  {journal} {\bibinfo  {journal} {Physics of Plasmas}\ }\textbf {\bibinfo {volume} {17}} (\bibinfo {year} {2010})}\BibitemShut {NoStop}%
\bibitem [{\citenamefont {Sovinec}\ \emph {et~al.}(2004)\citenamefont {Sovinec}, \citenamefont {Glasser}, \citenamefont {Gianakon}, \citenamefont {Barnes}, \citenamefont {Nebel}, \citenamefont {Kruger}, \citenamefont {Schnack}, \citenamefont {Plimpton}, \citenamefont {Tarditi}, \citenamefont {Chu} \emph {et~al.}}]{sovinec2004nimrod}%
  \BibitemOpen
  \bibfield  {author} {\bibinfo {author} {\bibfnamefont {C.~R.}\ \bibnamefont {Sovinec}}, \bibinfo {author} {\bibfnamefont {A.}~\bibnamefont {Glasser}}, \bibinfo {author} {\bibfnamefont {T.}~\bibnamefont {Gianakon}}, \bibinfo {author} {\bibfnamefont {D.}~\bibnamefont {Barnes}}, \bibinfo {author} {\bibfnamefont {R.}~\bibnamefont {Nebel}}, \bibinfo {author} {\bibfnamefont {S.}~\bibnamefont {Kruger}}, \bibinfo {author} {\bibfnamefont {D.}~\bibnamefont {Schnack}}, \bibinfo {author} {\bibfnamefont {S.}~\bibnamefont {Plimpton}}, \bibinfo {author} {\bibfnamefont {A.}~\bibnamefont {Tarditi}}, \bibinfo {author} {\bibfnamefont {M.}~\bibnamefont {Chu}},  \emph {et~al.},\ }\bibfield  {title} {\enquote {\bibinfo {title} {Nonlinear magnetohydrodynamics simulation using high-order finite elements},}\ }\href@noop {} {\bibfield  {journal} {\bibinfo  {journal} {Journal of Computational Physics}\ }\textbf {\bibinfo {volume} {195}},\ \bibinfo {pages} {355--386} (\bibinfo {year} {2004})}\BibitemShut {NoStop}%
\bibitem [{\citenamefont {Stone}\ \emph {et~al.}(2008{\natexlab{b}})\citenamefont {Stone}, \citenamefont {Gardiner}, \citenamefont {Teuben}, \citenamefont {Hawley},\ and\ \citenamefont {Simon}}]{stone2008athena}%
  \BibitemOpen
  \bibfield  {author} {\bibinfo {author} {\bibfnamefont {J.~M.}\ \bibnamefont {Stone}}, \bibinfo {author} {\bibfnamefont {T.~A.}\ \bibnamefont {Gardiner}}, \bibinfo {author} {\bibfnamefont {P.}~\bibnamefont {Teuben}}, \bibinfo {author} {\bibfnamefont {J.~F.}\ \bibnamefont {Hawley}}, \ and\ \bibinfo {author} {\bibfnamefont {J.~B.}\ \bibnamefont {Simon}},\ }\bibfield  {title} {\enquote {\bibinfo {title} {Athena: a new code for astrophysical mhd},}\ }\href@noop {} {\bibfield  {journal} {\bibinfo  {journal} {The Astrophysical Journal Supplement Series}\ }\textbf {\bibinfo {volume} {178}},\ \bibinfo {pages} {137} (\bibinfo {year} {2008}{\natexlab{b}})}\BibitemShut {NoStop}%
\bibitem [{\citenamefont {Hansen}\ \emph {et~al.}(2021)\citenamefont {Hansen}, \citenamefont {Reyes}, \citenamefont {Davies}, \citenamefont {Khiar}, \citenamefont {Adams}, \citenamefont {Armstrong}, \citenamefont {Farmakis}, \citenamefont {Lu}, \citenamefont {Michta}, \citenamefont {Moczulski} \emph {et~al.}}]{hansen2021extended}%
  \BibitemOpen
  \bibfield  {author} {\bibinfo {author} {\bibfnamefont {E.}~\bibnamefont {Hansen}}, \bibinfo {author} {\bibfnamefont {A.}~\bibnamefont {Reyes}}, \bibinfo {author} {\bibfnamefont {J.}~\bibnamefont {Davies}}, \bibinfo {author} {\bibfnamefont {B.}~\bibnamefont {Khiar}}, \bibinfo {author} {\bibfnamefont {M.}~\bibnamefont {Adams}}, \bibinfo {author} {\bibfnamefont {A.}~\bibnamefont {Armstrong}}, \bibinfo {author} {\bibfnamefont {P.}~\bibnamefont {Farmakis}}, \bibinfo {author} {\bibfnamefont {Y.}~\bibnamefont {Lu}}, \bibinfo {author} {\bibfnamefont {D.}~\bibnamefont {Michta}}, \bibinfo {author} {\bibfnamefont {K.}~\bibnamefont {Moczulski}},  \emph {et~al.},\ }\bibfield  {title} {\enquote {\bibinfo {title} {Extended magnetohydrodynamics in the flash code},}\ }in\ \href@noop {} {\emph {\bibinfo {booktitle} {APS Division of Plasma Physics Meeting Abstracts}}},\ Vol.\ \bibinfo {volume} {2021}\ (\bibinfo {year} {2021})\ pp.\ \bibinfo {pages} {CO08--007}\BibitemShut {NoStop}%
\bibitem [{\citenamefont {Navarro}, \citenamefont {Lora-Clavijo},\ and\ \citenamefont {Gonz{\'a}lez}(2017)}]{navarro2017magnus}%
  \BibitemOpen
  \bibfield  {author} {\bibinfo {author} {\bibfnamefont {A.}~\bibnamefont {Navarro}}, \bibinfo {author} {\bibfnamefont {F.}~\bibnamefont {Lora-Clavijo}}, \ and\ \bibinfo {author} {\bibfnamefont {G.~A.}\ \bibnamefont {Gonz{\'a}lez}},\ }\bibfield  {title} {\enquote {\bibinfo {title} {Magnus: a new resistive mhd code with heat flow terms},}\ }\href@noop {} {\bibfield  {journal} {\bibinfo  {journal} {The Astrophysical Journal}\ }\textbf {\bibinfo {volume} {844}},\ \bibinfo {pages} {57} (\bibinfo {year} {2017})}\BibitemShut {NoStop}%
\bibitem [{\citenamefont {Seyler}\ and\ \citenamefont {Martin}(2011)}]{seyler2011perseus}%
  \BibitemOpen
  \bibfield  {author} {\bibinfo {author} {\bibfnamefont {C.}~\bibnamefont {Seyler}}\ and\ \bibinfo {author} {\bibfnamefont {M.}~\bibnamefont {Martin}},\ }\bibfield  {title} {\enquote {\bibinfo {title} {Relaxation model for extended magnetohydrodynamics: Comparison to magnetohydrodynamics for dense z-pinches},}\ }\href@noop {} {\bibfield  {journal} {\bibinfo  {journal} {Physics of Plasmas}\ }\textbf {\bibinfo {volume} {18}} (\bibinfo {year} {2011})}\BibitemShut {NoStop}%
\bibitem [{\citenamefont {Ballinger}\ \emph {et~al.}(2021)\citenamefont {Ballinger}, \citenamefont {Kuang}, \citenamefont {Umansky}, \citenamefont {Brunner}, \citenamefont {Canik}, \citenamefont {Greenwald}, \citenamefont {Lore}, \citenamefont {LaBombard}, \citenamefont {Terry}, \citenamefont {Wigram} \emph {et~al.}}]{ballinger2021uedgeSPARC}%
  \BibitemOpen
  \bibfield  {author} {\bibinfo {author} {\bibfnamefont {S.}~\bibnamefont {Ballinger}}, \bibinfo {author} {\bibfnamefont {A.~Q.}\ \bibnamefont {Kuang}}, \bibinfo {author} {\bibfnamefont {M.}~\bibnamefont {Umansky}}, \bibinfo {author} {\bibfnamefont {D.}~\bibnamefont {Brunner}}, \bibinfo {author} {\bibfnamefont {J.~M.}\ \bibnamefont {Canik}}, \bibinfo {author} {\bibfnamefont {M.}~\bibnamefont {Greenwald}}, \bibinfo {author} {\bibfnamefont {J.~D.}\ \bibnamefont {Lore}}, \bibinfo {author} {\bibfnamefont {B.}~\bibnamefont {LaBombard}}, \bibinfo {author} {\bibfnamefont {J.~L.}\ \bibnamefont {Terry}}, \bibinfo {author} {\bibfnamefont {M.}~\bibnamefont {Wigram}},  \emph {et~al.},\ }\bibfield  {title} {\enquote {\bibinfo {title} {Simulation of the sparc plasma boundary with the uedge code},}\ }\href@noop {} {\bibfield  {journal} {\bibinfo  {journal} {Nuclear Fusion}\ }\textbf {\bibinfo {volume} {61}},\ \bibinfo {pages} {086014} (\bibinfo {year} {2021})}\BibitemShut {NoStop}%
\bibitem [{\citenamefont {Bonnin}\ \emph {et~al.}(2016)\citenamefont {Bonnin}, \citenamefont {Dekeyser}, \citenamefont {Pitts}, \citenamefont {Coster}, \citenamefont {Voskoboynikov},\ and\ \citenamefont {Wiesen}}]{bonnin2016SOLPSITER}%
  \BibitemOpen
  \bibfield  {author} {\bibinfo {author} {\bibfnamefont {X.}~\bibnamefont {Bonnin}}, \bibinfo {author} {\bibfnamefont {W.}~\bibnamefont {Dekeyser}}, \bibinfo {author} {\bibfnamefont {R.}~\bibnamefont {Pitts}}, \bibinfo {author} {\bibfnamefont {D.}~\bibnamefont {Coster}}, \bibinfo {author} {\bibfnamefont {S.}~\bibnamefont {Voskoboynikov}}, \ and\ \bibinfo {author} {\bibfnamefont {S.}~\bibnamefont {Wiesen}},\ }\bibfield  {title} {\enquote {\bibinfo {title} {Presentation of the new solps-iter code package for tokamak plasma edge modelling},}\ }\href@noop {} {\bibfield  {journal} {\bibinfo  {journal} {Plasma and Fusion Research}\ }\textbf {\bibinfo {volume} {11}},\ \bibinfo {pages} {1403102--1403102} (\bibinfo {year} {2016})}\BibitemShut {NoStop}%
\bibitem [{\citenamefont {Bufferand}\ \emph {et~al.}(2011)\citenamefont {Bufferand}, \citenamefont {Ciraolo}, \citenamefont {Isoardi}, \citenamefont {Chiavassa}, \citenamefont {Schwander}, \citenamefont {Serre}, \citenamefont {Fedorczak}, \citenamefont {Ghendrih},\ and\ \citenamefont {Tamain}}]{bufferand2011SOLEDGE}%
  \BibitemOpen
  \bibfield  {author} {\bibinfo {author} {\bibfnamefont {H.}~\bibnamefont {Bufferand}}, \bibinfo {author} {\bibfnamefont {G.}~\bibnamefont {Ciraolo}}, \bibinfo {author} {\bibfnamefont {L.}~\bibnamefont {Isoardi}}, \bibinfo {author} {\bibfnamefont {G.}~\bibnamefont {Chiavassa}}, \bibinfo {author} {\bibfnamefont {F.}~\bibnamefont {Schwander}}, \bibinfo {author} {\bibfnamefont {E.}~\bibnamefont {Serre}}, \bibinfo {author} {\bibfnamefont {N.}~\bibnamefont {Fedorczak}}, \bibinfo {author} {\bibfnamefont {P.}~\bibnamefont {Ghendrih}}, \ and\ \bibinfo {author} {\bibfnamefont {P.}~\bibnamefont {Tamain}},\ }\bibfield  {title} {\enquote {\bibinfo {title} {Applications of soledge-2d code to complex sol configurations and analysis of mach probe measurements},}\ }\href@noop {} {\bibfield  {journal} {\bibinfo  {journal} {Journal of Nuclear Materials}\ }\textbf {\bibinfo {volume} {415}},\ \bibinfo {pages} {S589--S592} (\bibinfo {year} {2011})}\BibitemShut {NoStop}%
\bibitem [{\citenamefont {Hillier}, \citenamefont {Takasao},\ and\ \citenamefont {Nakamura}(2016)}]{hillier2016slowshock}%
  \BibitemOpen
  \bibfield  {author} {\bibinfo {author} {\bibfnamefont {A.}~\bibnamefont {Hillier}}, \bibinfo {author} {\bibfnamefont {S.}~\bibnamefont {Takasao}}, \ and\ \bibinfo {author} {\bibfnamefont {N.}~\bibnamefont {Nakamura}},\ }\bibfield  {title} {\enquote {\bibinfo {title} {The formation and evolution of reconnection-driven, slow-mode shocks in a partially ionised plasma},}\ }\href@noop {} {\bibfield  {journal} {\bibinfo  {journal} {Astronomy \& Astrophysics}\ }\textbf {\bibinfo {volume} {591}},\ \bibinfo {pages} {A112} (\bibinfo {year} {2016})}\BibitemShut {NoStop}%
\bibitem [{\citenamefont {Reiter}, \citenamefont {Baelmans},\ and\ \citenamefont {Boerner}(2005)}]{reiter2005eirene}%
  \BibitemOpen
  \bibfield  {author} {\bibinfo {author} {\bibfnamefont {D.}~\bibnamefont {Reiter}}, \bibinfo {author} {\bibfnamefont {M.}~\bibnamefont {Baelmans}}, \ and\ \bibinfo {author} {\bibfnamefont {P.}~\bibnamefont {Boerner}},\ }\bibfield  {title} {\enquote {\bibinfo {title} {The eirene and b2-eirene codes},}\ }\href@noop {} {\bibfield  {journal} {\bibinfo  {journal} {Fusion science and technology}\ }\textbf {\bibinfo {volume} {47}},\ \bibinfo {pages} {172--186} (\bibinfo {year} {2005})}\BibitemShut {NoStop}%
\bibitem [{\citenamefont {Ciraolo}\ \emph {et~al.}(2019)\citenamefont {Ciraolo}, \citenamefont {Thin}, \citenamefont {Bufferand}, \citenamefont {Bucalossi}, \citenamefont {Fedorczak}, \citenamefont {Gunn}, \citenamefont {Pascal}, \citenamefont {Tamain}, \citenamefont {Gil}, \citenamefont {Gouin} \emph {et~al.}}]{ciraolo2019SOLEDGE_EIRENE}%
  \BibitemOpen
  \bibfield  {author} {\bibinfo {author} {\bibfnamefont {G.}~\bibnamefont {Ciraolo}}, \bibinfo {author} {\bibfnamefont {A.}~\bibnamefont {Thin}}, \bibinfo {author} {\bibfnamefont {H.}~\bibnamefont {Bufferand}}, \bibinfo {author} {\bibfnamefont {J.}~\bibnamefont {Bucalossi}}, \bibinfo {author} {\bibfnamefont {N.}~\bibnamefont {Fedorczak}}, \bibinfo {author} {\bibfnamefont {J.}~\bibnamefont {Gunn}}, \bibinfo {author} {\bibfnamefont {J.}~\bibnamefont {Pascal}}, \bibinfo {author} {\bibfnamefont {P.}~\bibnamefont {Tamain}}, \bibinfo {author} {\bibfnamefont {C.}~\bibnamefont {Gil}}, \bibinfo {author} {\bibfnamefont {A.}~\bibnamefont {Gouin}},  \emph {et~al.},\ }\bibfield  {title} {\enquote {\bibinfo {title} {First modeling of strongly radiating west plasmas with soledge-eirene},}\ }\href@noop {} {\bibfield  {journal} {\bibinfo  {journal} {Nuclear Materials and Energy}\ }\textbf {\bibinfo {volume} {20}},\ \bibinfo {pages} {100685} (\bibinfo {year} {2019})}\BibitemShut {NoStop}%
\bibitem [{\citenamefont {Borodin}\ \emph {et~al.}(2022)\citenamefont {Borodin}, \citenamefont {Schluck}, \citenamefont {Wiesen}, \citenamefont {Harting}, \citenamefont {Boerner}, \citenamefont {Brezinsek}, \citenamefont {Dekeyser}, \citenamefont {Carli}, \citenamefont {Blommaert}, \citenamefont {Van~Uytven} \emph {et~al.}}]{borodin2022edge}%
  \BibitemOpen
  \bibfield  {author} {\bibinfo {author} {\bibfnamefont {D.~V.}\ \bibnamefont {Borodin}}, \bibinfo {author} {\bibfnamefont {F.}~\bibnamefont {Schluck}}, \bibinfo {author} {\bibfnamefont {S.}~\bibnamefont {Wiesen}}, \bibinfo {author} {\bibfnamefont {D.}~\bibnamefont {Harting}}, \bibinfo {author} {\bibfnamefont {P.}~\bibnamefont {Boerner}}, \bibinfo {author} {\bibfnamefont {S.}~\bibnamefont {Brezinsek}}, \bibinfo {author} {\bibfnamefont {W.}~\bibnamefont {Dekeyser}}, \bibinfo {author} {\bibfnamefont {S.}~\bibnamefont {Carli}}, \bibinfo {author} {\bibfnamefont {M.}~\bibnamefont {Blommaert}}, \bibinfo {author} {\bibfnamefont {W.}~\bibnamefont {Van~Uytven}},  \emph {et~al.},\ }\bibfield  {title} {\enquote {\bibinfo {title} {Fluid, kinetic and hybrid approaches for neutral and trace ion edge transport modelling in fusion devices},}\ }\href@noop {} {\bibfield  {journal} {\bibinfo  {journal} {Nuclear Fusion}\ }\textbf {\bibinfo {volume} {62}},\ \bibinfo {pages} {086051} (\bibinfo {year} {2022})}\BibitemShut {NoStop}%
\bibitem [{\citenamefont {Krasheninnikov}, \citenamefont {Smolyakov},\ and\ \citenamefont {Kukushkin}(2020)}]{krasheninnikov2020edge}%
  \BibitemOpen
  \bibfield  {author} {\bibinfo {author} {\bibfnamefont {S.}~\bibnamefont {Krasheninnikov}}, \bibinfo {author} {\bibfnamefont {A.}~\bibnamefont {Smolyakov}}, \ and\ \bibinfo {author} {\bibfnamefont {A.}~\bibnamefont {Kukushkin}},\ }\href@noop {} {\emph {\bibinfo {title} {On the edge of magnetic fusion devices}}}\ (\bibinfo  {publisher} {Springer},\ \bibinfo {year} {2020})\BibitemShut {NoStop}%
\bibitem [{\citenamefont {Hillier}, \citenamefont {Snow},\ and\ \citenamefont {Luna}(2024)}]{hillier2024partially}%
  \BibitemOpen
  \bibfield  {author} {\bibinfo {author} {\bibfnamefont {A.~S.}\ \bibnamefont {Hillier}}, \bibinfo {author} {\bibfnamefont {B.}~\bibnamefont {Snow}}, \ and\ \bibinfo {author} {\bibfnamefont {M.}~\bibnamefont {Luna}},\ }\bibfield  {title} {\enquote {\bibinfo {title} {Partially ionized plasma of the solar atmosphere: recent advances and future pathways},}\ }\href@noop {} {\bibfield  {journal} {\bibinfo  {journal} {Philosophical Transactions of the Royal Society A}\ }\textbf {\bibinfo {volume} {382}},\ \bibinfo {pages} {20230230} (\bibinfo {year} {2024})}\BibitemShut {NoStop}%
\bibitem [{\citenamefont {Heinzel}, \citenamefont {Gun{\'a}r},\ and\ \citenamefont {Jej{\v{c}}i{\v{c}}}(2024)}]{heinzel2024partial}%
  \BibitemOpen
  \bibfield  {author} {\bibinfo {author} {\bibfnamefont {P.}~\bibnamefont {Heinzel}}, \bibinfo {author} {\bibfnamefont {S.}~\bibnamefont {Gun{\'a}r}}, \ and\ \bibinfo {author} {\bibfnamefont {S.}~\bibnamefont {Jej{\v{c}}i{\v{c}}}},\ }\bibfield  {title} {\enquote {\bibinfo {title} {Partial ionization of plasma in solar prominences},}\ }\href@noop {} {\bibfield  {journal} {\bibinfo  {journal} {Philosophical Transactions of the Royal Society A}\ }\textbf {\bibinfo {volume} {382}},\ \bibinfo {pages} {20230221} (\bibinfo {year} {2024})}\BibitemShut {NoStop}%
\bibitem [{\citenamefont {Soler}(2024)}]{soler2024magnetohydrodynamic}%
  \BibitemOpen
  \bibfield  {author} {\bibinfo {author} {\bibfnamefont {R.}~\bibnamefont {Soler}},\ }\bibfield  {title} {\enquote {\bibinfo {title} {Magnetohydrodynamic waves in the partially ionized solar plasma},}\ }\href@noop {} {\bibfield  {journal} {\bibinfo  {journal} {Philosophical Transactions of the Royal Society A}\ }\textbf {\bibinfo {volume} {382}},\ \bibinfo {pages} {20230223} (\bibinfo {year} {2024})}\BibitemShut {NoStop}%
\bibitem [{\citenamefont {Millmore}\ and\ \citenamefont {Nikiforakis}(2020)}]{millmore2020lightning}%
  \BibitemOpen
  \bibfield  {author} {\bibinfo {author} {\bibfnamefont {S.}~\bibnamefont {Millmore}}\ and\ \bibinfo {author} {\bibfnamefont {N.}~\bibnamefont {Nikiforakis}},\ }\bibfield  {title} {\enquote {\bibinfo {title} {Multi-physics simulations of lightning strike on elastoplastic substrates},}\ }\href@noop {} {\bibfield  {journal} {\bibinfo  {journal} {Journal of Computational Physics}\ }\textbf {\bibinfo {volume} {405}},\ \bibinfo {pages} {109142} (\bibinfo {year} {2020})}\BibitemShut {NoStop}%
\bibitem [{\citenamefont {Tr{\"a}uble}, \citenamefont {Millmore},\ and\ \citenamefont {Nikiforakis}(2021)}]{trauble2021improved}%
  \BibitemOpen
  \bibfield  {author} {\bibinfo {author} {\bibfnamefont {F.}~\bibnamefont {Tr{\"a}uble}}, \bibinfo {author} {\bibfnamefont {S.}~\bibnamefont {Millmore}}, \ and\ \bibinfo {author} {\bibfnamefont {N.}~\bibnamefont {Nikiforakis}},\ }\bibfield  {title} {\enquote {\bibinfo {title} {An improved equation of state for air plasma simulations},}\ }\href@noop {} {\bibfield  {journal} {\bibinfo  {journal} {Physics of Fluids}\ }\textbf {\bibinfo {volume} {33}} (\bibinfo {year} {2021})}\BibitemShut {NoStop}%
\bibitem [{\citenamefont {Wang}\ \emph {et~al.}(2023)\citenamefont {Wang}, \citenamefont {Ma}, \citenamefont {Wei}, \citenamefont {Wu},\ and\ \citenamefont {Huang}}]{wang2023lightning}%
  \BibitemOpen
  \bibfield  {author} {\bibinfo {author} {\bibfnamefont {F.}~\bibnamefont {Wang}}, \bibinfo {author} {\bibfnamefont {X.}~\bibnamefont {Ma}}, \bibinfo {author} {\bibfnamefont {Z.}~\bibnamefont {Wei}}, \bibinfo {author} {\bibfnamefont {Y.}~\bibnamefont {Wu}}, \ and\ \bibinfo {author} {\bibfnamefont {C.}~\bibnamefont {Huang}},\ }\bibfield  {title} {\enquote {\bibinfo {title} {Lightning damage of composite material driven by multi-physics coupling},}\ }\href@noop {} {\bibfield  {journal} {\bibinfo  {journal} {Composites Science and Technology}\ }\textbf {\bibinfo {volume} {233}},\ \bibinfo {pages} {109886} (\bibinfo {year} {2023})}\BibitemShut {NoStop}%
\bibitem [{\citenamefont {Xiao}\ and\ \citenamefont {Liu}(2023)}]{xiao2023flow_lightning}%
  \BibitemOpen
  \bibfield  {author} {\bibinfo {author} {\bibfnamefont {C.}~\bibnamefont {Xiao}}\ and\ \bibinfo {author} {\bibfnamefont {Y.}~\bibnamefont {Liu}},\ }\bibfield  {title} {\enquote {\bibinfo {title} {Influence of the aerodynamic flow on the dynamic characteristics of a lightning sweeping arc},}\ }\href@noop {} {\bibfield  {journal} {\bibinfo  {journal} {Frontiers in Astronomy and Space Sciences}\ }\textbf {\bibinfo {volume} {10}},\ \bibinfo {pages} {1083158} (\bibinfo {year} {2023})}\BibitemShut {NoStop}%
\bibitem [{\citenamefont {Auweter-Kurtz}, \citenamefont {Kurtz},\ and\ \citenamefont {Laure}(1996)}]{auweter1996generators_reentry}%
  \BibitemOpen
  \bibfield  {author} {\bibinfo {author} {\bibfnamefont {M.}~\bibnamefont {Auweter-Kurtz}}, \bibinfo {author} {\bibfnamefont {H.~L.}\ \bibnamefont {Kurtz}}, \ and\ \bibinfo {author} {\bibfnamefont {S.}~\bibnamefont {Laure}},\ }\bibfield  {title} {\enquote {\bibinfo {title} {Plasma generators for re-entry simulation},}\ }\href@noop {} {\bibfield  {journal} {\bibinfo  {journal} {Journal of Propulsion and Power}\ }\textbf {\bibinfo {volume} {12}},\ \bibinfo {pages} {1053--1061} (\bibinfo {year} {1996})}\BibitemShut {NoStop}%
\bibitem [{\citenamefont {Panesi}(2009)}]{panesi2009reentry}%
  \BibitemOpen
  \bibfield  {author} {\bibinfo {author} {\bibfnamefont {M.}~\bibnamefont {Panesi}},\ }\bibfield  {title} {\enquote {\bibinfo {title} {Physical models for nonequilibrium plasma flow simulations at high speed re-entry conditions},}\ }\href@noop {} {\bibfield  {journal} {\bibinfo  {journal} {Pise: Universit{\`a} di Pisa}\ } (\bibinfo {year} {2009})}\BibitemShut {NoStop}%
\bibitem [{\citenamefont {Macheret}(2008)}]{macheret2008weakIonize}%
  \BibitemOpen
  \bibfield  {author} {\bibinfo {author} {\bibfnamefont {S.}~\bibnamefont {Macheret}},\ }\bibfield  {title} {\enquote {\bibinfo {title} {Introduction: Weakly ionized plasmas for propulsion applications},}\ }\href@noop {} {\bibfield  {journal} {\bibinfo  {journal} {Journal of Propulsion and Power}\ }\textbf {\bibinfo {volume} {24}},\ \bibinfo {pages} {898--899} (\bibinfo {year} {2008})}\BibitemShut {NoStop}%
\bibitem [{\citenamefont {Stotler}\ and\ \citenamefont {Karney}(1994)}]{stotler1994degas2}%
  \BibitemOpen
  \bibfield  {author} {\bibinfo {author} {\bibfnamefont {D.}~\bibnamefont {Stotler}}\ and\ \bibinfo {author} {\bibfnamefont {C.}~\bibnamefont {Karney}},\ }\bibfield  {title} {\enquote {\bibinfo {title} {Neutral gas transport modeling with degas 2},}\ }\href@noop {} {\bibfield  {journal} {\bibinfo  {journal} {Contributions to Plasma Physics}\ }\textbf {\bibinfo {volume} {34}},\ \bibinfo {pages} {392--397} (\bibinfo {year} {1994})}\BibitemShut {NoStop}%
\bibitem [{\citenamefont {Braginskii}(1965)}]{braginskii1965}%
  \BibitemOpen
  \bibfield  {author} {\bibinfo {author} {\bibfnamefont {S.}~\bibnamefont {Braginskii}},\ }\bibfield  {title} {\enquote {\bibinfo {title} {Transport processes in a plasma},}\ }\href@noop {} {\bibfield  {journal} {\bibinfo  {journal} {Reviews of plasma physics}\ }\textbf {\bibinfo {volume} {1}},\ \bibinfo {pages} {205} (\bibinfo {year} {1965})}\BibitemShut {NoStop}%
\bibitem [{\citenamefont {Riemann}(1991)}]{riemann1991bohm}%
  \BibitemOpen
  \bibfield  {author} {\bibinfo {author} {\bibfnamefont {K.-U.}\ \bibnamefont {Riemann}},\ }\bibfield  {title} {\enquote {\bibinfo {title} {The bohm criterion and sheath formation},}\ }\href@noop {} {\bibfield  {journal} {\bibinfo  {journal} {Journal of Physics D: Applied Physics}\ }\textbf {\bibinfo {volume} {24}},\ \bibinfo {pages} {493} (\bibinfo {year} {1991})}\BibitemShut {NoStop}%
\bibitem [{\citenamefont {Stotler}\ \emph {et~al.}(2003)\citenamefont {Stotler}, \citenamefont {LaBombard}, \citenamefont {Terry},\ and\ \citenamefont {Zweben}}]{stotler2003puffing}%
  \BibitemOpen
  \bibfield  {author} {\bibinfo {author} {\bibfnamefont {D.}~\bibnamefont {Stotler}}, \bibinfo {author} {\bibfnamefont {B.}~\bibnamefont {LaBombard}}, \bibinfo {author} {\bibfnamefont {J.}~\bibnamefont {Terry}}, \ and\ \bibinfo {author} {\bibfnamefont {S.}~\bibnamefont {Zweben}},\ }\bibfield  {title} {\enquote {\bibinfo {title} {Neutral transport simulations of gas puff imaging experiments},}\ }\href@noop {} {\bibfield  {journal} {\bibinfo  {journal} {Journal of nuclear materials}\ }\textbf {\bibinfo {volume} {313}},\ \bibinfo {pages} {1066--1070} (\bibinfo {year} {2003})}\BibitemShut {NoStop}%
\bibitem [{\citenamefont {Toro}(2013)}]{toro2013riemann}%
  \BibitemOpen
  \bibfield  {author} {\bibinfo {author} {\bibfnamefont {E.}~\bibnamefont {Toro}},\ }\href@noop {} {\emph {\bibinfo {title} {Riemann solvers and numerical methods for fluid dynamics: a practical introduction}}}\ (\bibinfo  {publisher} {Springer Science \& Business Media},\ \bibinfo {year} {2013})\BibitemShut {NoStop}%
\bibitem [{\citenamefont {Scoggins}\ \emph {et~al.}(2020)\citenamefont {Scoggins}, \citenamefont {Leroy}, \citenamefont {Bellas-Chatzigeorgis}, \citenamefont {Dias},\ and\ \citenamefont {Magin}}]{scoggins2020mutation++}%
  \BibitemOpen
  \bibfield  {author} {\bibinfo {author} {\bibfnamefont {J.~B.}\ \bibnamefont {Scoggins}}, \bibinfo {author} {\bibfnamefont {V.}~\bibnamefont {Leroy}}, \bibinfo {author} {\bibfnamefont {G.}~\bibnamefont {Bellas-Chatzigeorgis}}, \bibinfo {author} {\bibfnamefont {B.}~\bibnamefont {Dias}}, \ and\ \bibinfo {author} {\bibfnamefont {T.~E.}\ \bibnamefont {Magin}},\ }\bibfield  {title} {\enquote {\bibinfo {title} {Mutation++: Multicomponent thermodynamic and transport properties for ionized gases in c++},}\ }\href@noop {} {\bibfield  {journal} {\bibinfo  {journal} {SoftwareX}\ }\textbf {\bibinfo {volume} {12}},\ \bibinfo {pages} {100575} (\bibinfo {year} {2020})}\BibitemShut {NoStop}%
\bibitem [{\citenamefont {Gordon}\ and\ \citenamefont {McBride}(1994)}]{gordon1994CEA}%
  \BibitemOpen
  \bibfield  {author} {\bibinfo {author} {\bibfnamefont {S.}~\bibnamefont {Gordon}}\ and\ \bibinfo {author} {\bibfnamefont {B.~J.}\ \bibnamefont {McBride}},\ }\href@noop {} {\enquote {\bibinfo {title} {Computer program for calculation of complex chemical equilibrium compositions and applications. part 1: Analysis},}\ }\bibinfo {type} {Tech. Rep.}\ (\bibinfo {year} {1994})\BibitemShut {NoStop}%
\bibitem [{\citenamefont {Ern}\ and\ \citenamefont {Giovangigli}(2004)}]{ern2004eglib}%
  \BibitemOpen
  \bibfield  {author} {\bibinfo {author} {\bibfnamefont {A.}~\bibnamefont {Ern}}\ and\ \bibinfo {author} {\bibfnamefont {V.}~\bibnamefont {Giovangigli}},\ }\bibfield  {title} {\enquote {\bibinfo {title} {Eglib: A general-purpose fortran library for multicomponent transport property evaluation},}\ }\href@noop {} {\bibfield  {journal} {\bibinfo  {journal} {Manual of EGlib version}\ }\textbf {\bibinfo {volume} {3}},\ \bibinfo {pages} {12} (\bibinfo {year} {2004})}\BibitemShut {NoStop}%
\bibitem [{\citenamefont {Kee}(1999)}]{kee1999chemkin}%
  \BibitemOpen
  \bibfield  {author} {\bibinfo {author} {\bibfnamefont {R.}~\bibnamefont {Kee}},\ }\bibfield  {title} {\enquote {\bibinfo {title} {Chemkin collection},}\ }\href@noop {} {\bibfield  {journal} {\bibinfo  {journal} {Release}\ }\textbf {\bibinfo {volume} {3}} (\bibinfo {year} {1999})}\BibitemShut {NoStop}%
\bibitem [{\citenamefont {Pope}(2003)}]{pope2003gibbsCont}%
  \BibitemOpen
  \bibfield  {author} {\bibinfo {author} {\bibfnamefont {S.~B.}\ \bibnamefont {Pope}},\ }\bibfield  {title} {\enquote {\bibinfo {title} {The computation of constrained and unconstrained equilibrium compositions of ideal gas mixtures using gibbs function continuation},}\ }\href@noop {} {\bibfield  {journal} {\bibinfo  {journal} {Cornell University Report FDA}\ ,\ \bibinfo {pages} {03--02}} (\bibinfo {year} {2003})}\BibitemShut {NoStop}%
\bibitem [{\citenamefont {McBride}(2002)}]{mcbride2002nasa}%
  \BibitemOpen
  \bibfield  {author} {\bibinfo {author} {\bibfnamefont {B.~J.}\ \bibnamefont {McBride}},\ }\href@noop {} {\emph {\bibinfo {title} {NASA Glenn coefficients for calculating thermodynamic properties of individual species}}}\ (\bibinfo  {publisher} {National Aeronautics and Space Administration, John H. Glenn Research Center~…},\ \bibinfo {year} {2002})\BibitemShut {NoStop}%
\bibitem [{\citenamefont {Gordon}\ and\ \citenamefont {McBride}(1999)}]{gordon1999thermodynamic}%
  \BibitemOpen
  \bibfield  {author} {\bibinfo {author} {\bibfnamefont {S.}~\bibnamefont {Gordon}}\ and\ \bibinfo {author} {\bibfnamefont {B.~J.}\ \bibnamefont {McBride}},\ }\href@noop {} {\enquote {\bibinfo {title} {Thermodynamic data to 20,000 k for monatomic gases},}\ }\bibinfo {type} {Tech. Rep.}\ (\bibinfo {year} {1999})\BibitemShut {NoStop}%
\bibitem [{\citenamefont {Strang}(1968)}]{strang1968construction}%
  \BibitemOpen
  \bibfield  {author} {\bibinfo {author} {\bibfnamefont {G.}~\bibnamefont {Strang}},\ }\bibfield  {title} {\enquote {\bibinfo {title} {On the construction and comparison of difference schemes},}\ }\href@noop {} {\bibfield  {journal} {\bibinfo  {journal} {SIAM journal on numerical analysis}\ }\textbf {\bibinfo {volume} {5}},\ \bibinfo {pages} {506--517} (\bibinfo {year} {1968})}\BibitemShut {NoStop}%
\bibitem [{\citenamefont {De~Moura}\ and\ \citenamefont {Kubrusly}(2013)}]{de2013cfl}%
  \BibitemOpen
  \bibfield  {author} {\bibinfo {author} {\bibfnamefont {C.~A.}\ \bibnamefont {De~Moura}}\ and\ \bibinfo {author} {\bibfnamefont {C.~S.}\ \bibnamefont {Kubrusly}},\ }\bibfield  {title} {\enquote {\bibinfo {title} {The courant--friedrichs--lewy (cfl) condition},}\ }\href@noop {} {\bibfield  {journal} {\bibinfo  {journal} {AMC}\ }\textbf {\bibinfo {volume} {10}},\ \bibinfo {pages} {45--90} (\bibinfo {year} {2013})}\BibitemShut {NoStop}%
\bibitem [{\citenamefont {Crank}\ and\ \citenamefont {Nicolson}(1947)}]{crankNicolson}%
  \BibitemOpen
  \bibfield  {author} {\bibinfo {author} {\bibfnamefont {J.}~\bibnamefont {Crank}}\ and\ \bibinfo {author} {\bibfnamefont {P.}~\bibnamefont {Nicolson}},\ }\bibfield  {title} {\enquote {\bibinfo {title} {A practical method for numerical evaluation of solutions of partial differential equations of the heat-conduction type},}\ }in\ \href@noop {} {\emph {\bibinfo {booktitle} {Mathematical proceedings of the Cambridge philosophical society}}},\ Vol.~\bibinfo {volume} {43}\ (\bibinfo {organization} {Cambridge University Press},\ \bibinfo {year} {1947})\ pp.\ \bibinfo {pages} {50--67}\BibitemShut {NoStop}%
\bibitem [{\citenamefont {Butcher}(2016)}]{butcher2016numerical}%
  \BibitemOpen
  \bibfield  {author} {\bibinfo {author} {\bibfnamefont {J.~C.}\ \bibnamefont {Butcher}},\ }\href@noop {} {\emph {\bibinfo {title} {Numerical methods for ordinary differential equations}}}\ (\bibinfo  {publisher} {John Wiley \& Sons},\ \bibinfo {year} {2016})\BibitemShut {NoStop}%
\bibitem [{\citenamefont {{Dedner}}\ \emph {et~al.}(2002)\citenamefont {{Dedner}}, \citenamefont {{Kemm}}, \citenamefont {{Kr{\"o}ner}}, \citenamefont {{Munz}}, \citenamefont {{Schnitzer}},\ and\ \citenamefont {{Wesenberg}}}]{Dedner2002}%
  \BibitemOpen
  \bibfield  {author} {\bibinfo {author} {\bibfnamefont {A.}~\bibnamefont {{Dedner}}}, \bibinfo {author} {\bibfnamefont {F.}~\bibnamefont {{Kemm}}}, \bibinfo {author} {\bibfnamefont {D.}~\bibnamefont {{Kr{\"o}ner}}}, \bibinfo {author} {\bibfnamefont {C.~D.}\ \bibnamefont {{Munz}}}, \bibinfo {author} {\bibfnamefont {T.}~\bibnamefont {{Schnitzer}}}, \ and\ \bibinfo {author} {\bibfnamefont {M.}~\bibnamefont {{Wesenberg}}},\ }\bibfield  {title} {\enquote {\bibinfo {title} {{Hyperbolic Divergence Cleaning for the MHD Equations}},}\ }\href {\doibase 10.1006/jcph.2001.6961} {\bibfield  {journal} {\bibinfo  {journal} {Journal of Computational Physics}\ }\textbf {\bibinfo {volume} {175}},\ \bibinfo {pages} {645--673} (\bibinfo {year} {2002})}\BibitemShut {NoStop}%
\bibitem [{\citenamefont {Brio}\ and\ \citenamefont {Wu}(1988)}]{brioWu1988}%
  \BibitemOpen
  \bibfield  {author} {\bibinfo {author} {\bibfnamefont {M.}~\bibnamefont {Brio}}\ and\ \bibinfo {author} {\bibfnamefont {C.~C.}\ \bibnamefont {Wu}},\ }\bibfield  {title} {\enquote {\bibinfo {title} {An upwind differencing scheme for the equations of ideal magnetohydrodynamics},}\ }\href@noop {} {\bibfield  {journal} {\bibinfo  {journal} {Journal of computational physics}\ }\textbf {\bibinfo {volume} {75}},\ \bibinfo {pages} {400--422} (\bibinfo {year} {1988})}\BibitemShut {NoStop}%
\bibitem [{\citenamefont {Farmer}, \citenamefont {Ockendon},\ and\ \citenamefont {Ockendon}(2021)}]{farmer2021_woodspeed}%
  \BibitemOpen
  \bibfield  {author} {\bibinfo {author} {\bibfnamefont {C.}~\bibnamefont {Farmer}}, \bibinfo {author} {\bibfnamefont {H.}~\bibnamefont {Ockendon}}, \ and\ \bibinfo {author} {\bibfnamefont {J.}~\bibnamefont {Ockendon}},\ }\bibfield  {title} {\enquote {\bibinfo {title} {One-dimensional acoustic waves in air/water mixtures},}\ }\href@noop {} {\bibfield  {journal} {\bibinfo  {journal} {Wave Motion}\ }\textbf {\bibinfo {volume} {106}},\ \bibinfo {pages} {102798} (\bibinfo {year} {2021})}\BibitemShut {NoStop}%
\bibitem [{\citenamefont {Kieffer}(1977)}]{kieffer1977sound}%
  \BibitemOpen
  \bibfield  {author} {\bibinfo {author} {\bibfnamefont {S.~W.}\ \bibnamefont {Kieffer}},\ }\bibfield  {title} {\enquote {\bibinfo {title} {Sound speed in liquid-gas mixtures: Water-air and water-steam},}\ }\href@noop {} {\bibfield  {journal} {\bibinfo  {journal} {Journal of Geophysical research}\ }\textbf {\bibinfo {volume} {82}},\ \bibinfo {pages} {2895--2904} (\bibinfo {year} {1977})}\BibitemShut {NoStop}%
\bibitem [{\citenamefont {Needham}(2010)}]{needham2010blast}%
  \BibitemOpen
  \bibfield  {author} {\bibinfo {author} {\bibfnamefont {C.~E.}\ \bibnamefont {Needham}},\ }\href@noop {} {\emph {\bibinfo {title} {Blast waves}}},\ Vol.\ \bibinfo {volume} {402}\ (\bibinfo  {publisher} {Springer},\ \bibinfo {year} {2010})\BibitemShut {NoStop}%
\bibitem [{\citenamefont {Parra}(2018)}]{parra2018resistive}%
  \BibitemOpen
  \bibfield  {author} {\bibinfo {author} {\bibfnamefont {F.~I.}\ \bibnamefont {Parra}},\ }\href@noop {} {\enquote {\bibinfo {title} {Resistive mhd, reconnection, and resistive tearing modes},}\ } (\bibinfo {year} {2018})\BibitemShut {NoStop}%
\bibitem [{\citenamefont {Banerjee}\ \emph {et~al.}(2024)\citenamefont {Banerjee}, \citenamefont {Boyle}, \citenamefont {Maan}, \citenamefont {Ferraro}, \citenamefont {Wilkie}, \citenamefont {Majeski}, \citenamefont {Podesta}, \citenamefont {Bell}, \citenamefont {Hansen}, \citenamefont {Capecchi} \emph {et~al.}}]{banerjee2024neutral_tearing_mode}%
  \BibitemOpen
  \bibfield  {author} {\bibinfo {author} {\bibfnamefont {S.}~\bibnamefont {Banerjee}}, \bibinfo {author} {\bibfnamefont {D.}~\bibnamefont {Boyle}}, \bibinfo {author} {\bibfnamefont {A.}~\bibnamefont {Maan}}, \bibinfo {author} {\bibfnamefont {N.}~\bibnamefont {Ferraro}}, \bibinfo {author} {\bibfnamefont {G.}~\bibnamefont {Wilkie}}, \bibinfo {author} {\bibfnamefont {R.}~\bibnamefont {Majeski}}, \bibinfo {author} {\bibfnamefont {M.}~\bibnamefont {Podesta}}, \bibinfo {author} {\bibfnamefont {R.}~\bibnamefont {Bell}}, \bibinfo {author} {\bibfnamefont {C.}~\bibnamefont {Hansen}}, \bibinfo {author} {\bibfnamefont {W.}~\bibnamefont {Capecchi}},  \emph {et~al.},\ }\bibfield  {title} {\enquote {\bibinfo {title} {Investigating the role of edge neutrals in exciting tearing mode activity and achieving flat temperature profiles in ltx-$\beta$},}\ }\href@noop {} {\bibfield  {journal} {\bibinfo  {journal} {Nuclear Fusion}\ }\textbf {\bibinfo {volume} {64}},\ \bibinfo {pages} {046026} (\bibinfo {year} {2024})}\BibitemShut
  {NoStop}%
\bibitem [{\citenamefont {Zheng}\ and\ \citenamefont {Qiu}(2024)}]{zheng2024MGEoS}%
  \BibitemOpen
  \bibfield  {author} {\bibinfo {author} {\bibfnamefont {F.}~\bibnamefont {Zheng}}\ and\ \bibinfo {author} {\bibfnamefont {J.}~\bibnamefont {Qiu}},\ }\bibfield  {title} {\enquote {\bibinfo {title} {High-order finite volume method for solving compressible multicomponent flows with mie--gr{\"u}neisen equation of state},}\ }\href@noop {} {\bibfield  {journal} {\bibinfo  {journal} {Computers \& Fluids}\ }\textbf {\bibinfo {volume} {284}},\ \bibinfo {pages} {106424} (\bibinfo {year} {2024})}\BibitemShut {NoStop}%
\bibitem [{\citenamefont {Glatzmaier}()}]{UCAstroNotes}%
  \BibitemOpen
  \bibfield  {author} {\bibinfo {author} {\bibfnamefont {G.~A.}\ \bibnamefont {Glatzmaier}},\ }\href@noop {} {\enquote {\bibinfo {title} {Internal energy and radiative transfer},}\ }\bibinfo {howpublished} {Astronomy 112: The Physics of Stars}\BibitemShut {NoStop}%
\bibitem [{\citenamefont {Krommes}(2019)}]{krommes2019CLog}%
  \BibitemOpen
  \bibfield  {author} {\bibinfo {author} {\bibfnamefont {J.}~\bibnamefont {Krommes}},\ }\bibfield  {title} {\enquote {\bibinfo {title} {An introduction to the physics of the coulomb logarithm, with emphasis on quantum-mechanical effects},}\ }\href@noop {} {\bibfield  {journal} {\bibinfo  {journal} {Journal of Plasma Physics}\ }\textbf {\bibinfo {volume} {85}} (\bibinfo {year} {2019})}\BibitemShut {NoStop}%
\bibitem [{\citenamefont {Spitzer}(1956)}]{spitzer1956physics}%
  \BibitemOpen
  \bibfield  {author} {\bibinfo {author} {\bibfnamefont {L.}~\bibnamefont {Spitzer}},\ }\href@noop {} {\emph {\bibinfo {title} {Physics of fully ionized gases}}}\ (\bibinfo  {publisher} {Courier Corporation},\ \bibinfo {year} {1956})\BibitemShut {NoStop}%
\bibitem [{\citenamefont {Vranjes}\ and\ \citenamefont {Krstic}(2013)}]{vranjes2013collisions}%
  \BibitemOpen
  \bibfield  {author} {\bibinfo {author} {\bibfnamefont {J.}~\bibnamefont {Vranjes}}\ and\ \bibinfo {author} {\bibfnamefont {P.}~\bibnamefont {Krstic}},\ }\bibfield  {title} {\enquote {\bibinfo {title} {Collisions, magnetization, and transport coefficients in the lower solar atmosphere},}\ }\href@noop {} {\bibfield  {journal} {\bibinfo  {journal} {Astronomy \& Astrophysics}\ }\textbf {\bibinfo {volume} {554}},\ \bibinfo {pages} {A22} (\bibinfo {year} {2013})}\BibitemShut {NoStop}%
\bibitem [{\citenamefont {Pandey}\ and\ \citenamefont {Wardle}(2008)}]{pandey2008hall}%
  \BibitemOpen
  \bibfield  {author} {\bibinfo {author} {\bibfnamefont {B.}~\bibnamefont {Pandey}}\ and\ \bibinfo {author} {\bibfnamefont {M.}~\bibnamefont {Wardle}},\ }\bibfield  {title} {\enquote {\bibinfo {title} {Hall magnetohydrodynamics of partially ionized plasmas},}\ }\href@noop {} {\bibfield  {journal} {\bibinfo  {journal} {Monthly Notices of the Royal Astronomical Society}\ }\textbf {\bibinfo {volume} {385}},\ \bibinfo {pages} {2269--2278} (\bibinfo {year} {2008})}\BibitemShut {NoStop}%
\bibitem [{\citenamefont {Ramshaw}\ and\ \citenamefont {Chang}(1991)}]{ramshaw1991ambipolar}%
  \BibitemOpen
  \bibfield  {author} {\bibinfo {author} {\bibfnamefont {J.}~\bibnamefont {Ramshaw}}\ and\ \bibinfo {author} {\bibfnamefont {C.}~\bibnamefont {Chang}},\ }\bibfield  {title} {\enquote {\bibinfo {title} {Ambipolar diffusion in multicomponent plasmas},}\ }\href@noop {} {\bibfield  {journal} {\bibinfo  {journal} {Plasma chemistry and plasma processing}\ }\textbf {\bibinfo {volume} {11}},\ \bibinfo {pages} {395--402} (\bibinfo {year} {1991})}\BibitemShut {NoStop}%
\bibitem [{\citenamefont {Khomenko}\ \emph {et~al.}(2014)\citenamefont {Khomenko}, \citenamefont {Collados}, \citenamefont {Diaz},\ and\ \citenamefont {Vitas}}]{khomenko2014fluid}%
  \BibitemOpen
  \bibfield  {author} {\bibinfo {author} {\bibfnamefont {E.}~\bibnamefont {Khomenko}}, \bibinfo {author} {\bibfnamefont {M.}~\bibnamefont {Collados}}, \bibinfo {author} {\bibfnamefont {A.}~\bibnamefont {Diaz}}, \ and\ \bibinfo {author} {\bibfnamefont {N.}~\bibnamefont {Vitas}},\ }\bibfield  {title} {\enquote {\bibinfo {title} {Fluid description of multi-component solar partially ionized plasma},}\ }\href@noop {} {\bibfield  {journal} {\bibinfo  {journal} {Physics of Plasmas}\ }\textbf {\bibinfo {volume} {21}} (\bibinfo {year} {2014})}\BibitemShut {NoStop}%
\bibitem [{\citenamefont {Aymar}, \citenamefont {Barabaschi},\ and\ \citenamefont {Shimomura}(2002)}]{aymar2002iter}%
  \BibitemOpen
  \bibfield  {author} {\bibinfo {author} {\bibfnamefont {R.}~\bibnamefont {Aymar}}, \bibinfo {author} {\bibfnamefont {P.}~\bibnamefont {Barabaschi}}, \ and\ \bibinfo {author} {\bibfnamefont {Y.}~\bibnamefont {Shimomura}},\ }\bibfield  {title} {\enquote {\bibinfo {title} {The iter design},}\ }\href@noop {} {\bibfield  {journal} {\bibinfo  {journal} {Plasma physics and controlled fusion}\ }\textbf {\bibinfo {volume} {44}},\ \bibinfo {pages} {519} (\bibinfo {year} {2002})}\BibitemShut {NoStop}%
\bibitem [{\citenamefont {Sod}(1978)}]{sod1978survey}%
  \BibitemOpen
  \bibfield  {author} {\bibinfo {author} {\bibfnamefont {G.~A.}\ \bibnamefont {Sod}},\ }\bibfield  {title} {\enquote {\bibinfo {title} {A survey of several finite difference methods for systems of nonlinear hyperbolic conservation laws},}\ }\href@noop {} {\bibfield  {journal} {\bibinfo  {journal} {Journal of computational physics}\ }\textbf {\bibinfo {volume} {27}},\ \bibinfo {pages} {1--31} (\bibinfo {year} {1978})}\BibitemShut {NoStop}%
\bibitem [{\citenamefont {Jiang}, \citenamefont {Fang},\ and\ \citenamefont {Chen}(2012)}]{jiang2012new}%
  \BibitemOpen
  \bibfield  {author} {\bibinfo {author} {\bibfnamefont {R.-L.}\ \bibnamefont {Jiang}}, \bibinfo {author} {\bibfnamefont {C.}~\bibnamefont {Fang}}, \ and\ \bibinfo {author} {\bibfnamefont {P.-F.}\ \bibnamefont {Chen}},\ }\bibfield  {title} {\enquote {\bibinfo {title} {A new mhd code with adaptive mesh refinement and parallelization for astrophysics},}\ }\href@noop {} {\bibfield  {journal} {\bibinfo  {journal} {Computer Physics Communications}\ }\textbf {\bibinfo {volume} {183}},\ \bibinfo {pages} {1617--1633} (\bibinfo {year} {2012})}\BibitemShut {NoStop}%
\end{thebibliography}%

\end{document}